\documentclass[10pt]{amsart}
\usepackage{amsmath,amssymb,amsthm} 
\usepackage[T1]{fontenc}
\usepackage[utf8]{inputenc}
\usepackage{lmodern}
\usepackage{graphicx}
\usepackage{subcaption}
\usepackage{chronology}
\usepackage{csquotes}
\usepackage[english]{babel}
\usepackage{url}
\usepackage{amsthm}
\usepackage{amsmath}
\usepackage{amsfonts}
\usepackage{amssymb}
\usepackage[shortlabels]{enumitem}
\usepackage{listings}
\usepackage{xcolor}
\usepackage{bbm}
\usepackage{rotating}
\usepackage{tikz}
\usepackage{placeins}
\usepackage{amsaddr}
\usepackage[round, compress]{natbib}
\usepackage{hyperref}
\usepackage{comment}
\usepackage[margin=1in]{geometry} 
\usepackage{hyperref}
\usepackage{hyperref}
\usepackage{mathtools}
\usepackage{xcolor}
\usepackage{booktabs}
\usepackage{arydshln}
\usepackage{makecell}
\usepackage{amsaddr} 
\usepackage{hyperref} 

\usepackage{lmodern}
\usepackage[margin=1in]{geometry}

\usepackage{hyperref} 

\numberwithin{equation}{section}
\setcitestyle{numbers}

\definecolor{measuredColor}{rgb}{0.094, 0.631, 0.466}
\definecolor{estimatedColor}{rgb}{0.906, 0.623, 0.149}
\definecolor{calibratedColor}{rgb}{0.043, 0.454, 0.71}
\definecolor{fixedColor}{rgb}{0.835, 0.384, 0.16}

\newcommand*{\affmark}[1]{\textsuperscript{#1}}

\usepackage{caption}
\newcommand{\subtitle}[1]{%
	\posttitle{%
		\par\end{center}
	\begin{center}\large#1\end{center}
	\vskip0.1em}}%

\makeatletter
\def\subsection{\@startsection{subsection}{3}%
  \z@{.5\linespacing\@plus.7\linespacing}{.1\linespacing}%
  {\normalfont\scshape\centering}}
\makeatother

\makeatletter
\def\subsubsection{\@startsection{subsubsection}{3}%
  \z@{.5\linespacing\@plus.7\linespacing}{.1\linespacing}%
  {\normalfont\itshape}}
\makeatother

\begin{document}

\title[A Global Systems Perspective on Food Demand, Deforestation and Agricultural Sustainability]{A Global Systems Perspective on Food Demand, Deforestation and Agricultural Sustainability}

\author{Elia Moretti\affmark{1,2},
        Michel Loreau\affmark{3,4} \and
        Michael Benzaquen\affmark{1,2,5}}

\maketitle

\begin{center}
\small
\affmark{1}Chair of Econophysics and Complex Systems, École polytechnique, 91128 Palaiseau Cedex, France.\\
\affmark{2}LadHyX, CNRS, École polytechnique, Institut Polytechnique de Paris, 91120 Palaiseau, France.\\

\affmark{3}Station d'écologie théorique et expérimentale, CNRS, 09200 Moulis, France.\\
\affmark{4}Institute of Ecology, College of Urban and Environmental Sciences, Peking University, Beijing 100871, China.\\
\affmark{5}Capital Fund Management, 23 rue de l'Université, 75007 Paris, France.
\end{center}

\begin{abstract}

Feeding a larger and wealthier global population without transgressing ecological limits is increasingly challenging, as rising food demand—especially for animal products—intensifies pressure on ecosystems, accelerates deforestation, and erodes biodiversity and soil health.
We develop a stylized, spatially explicit global model that links exogenous food-demand trajectories to crop and livestock production, land conversion, and feedbacks from ecosystem integrity that, in turn, shape future yields and land needs. Calibrated to post-1960 trends in population, income, yields, input use, and land use, the model reproduces the joint rise of crop and meat demand and the associated expansion and intensification of agriculture. We use it to compare business-as-usual, supply-side, demand-side, and mixed-policy scenarios. Three results stand out. First, productivity-oriented supply-side measures (e.g. reduced chemical inputs, organic conversion, lower livestock density) often trigger compensatory land expansion that undermines ecological gains—so that supply-side action alone cannot halt deforestation or widespread degradation. Second, demand-side change, particularly reduced meat consumption, consistently relieves both intensification and expansion pressures; in our simulations, only substantial demand reductions (on the order of ~40\% of projected excess demand by 2100) deliver simultaneous increases in forest area and declines in degraded land. Third, integrated policy portfolios that jointly constrain land conversion, temper input intensification, and curb demand outperform any single lever. Together, these findings clarify the system-level trade-offs that frustrate piecemeal interventions and identify the policy combinations most likely to keep global food provision within ecological limits.


\end{abstract}


\section{Introduction}

Feeding a growing and wealthier population in the 21st century presents one of the biggest challenges to planetary sustainability. Rapid income growth and global population expansion are driving up the total demand for crops and for animal products, thereby escalating the pressures on the natural ecosystem. This increasing food demand often comes at the expense of forests, semi-natural ecosystems, and long-term soil health, fueling deforestation, biodiversity loss, and land degradation at unprecedented scales~\cite{Foley2011, Tilman2011, Pendrill2022}. As a result, reconciling food security with the safe operating space for humanity—maintaining ecological integrity while meeting human needs—has emerged as a global policy and scientific imperative~\cite{Rockstrom2009,Willett2019, loreau2010populations}.

A central question in this debate is whether ongoing increases in agricultural productivity, alongside the adoption of more sustainable farming practices—such as organic and conservation agriculture, reductions in chemical inputs, or lower livestock densities—can suffice to meet rising demand while preserving ecosystems~\cite{Reganold2016, Seufert2012}. Or, conversely, is a fundamental transformation of demand, especially through dietary shifts and coordinated, system-wide policy measures, necessary to avoid crossing more critical environmental thresholds? In light of clear trade-offs between food production, land use, and environmental services, understanding the boundaries and synergies of supply-side versus demand-side strategies remains an open challenge~\cite{Garnett2013,Bajzelj2014}.

The literature tackling these questions spans empirical, scenario-based, and theoretical approaches. Global integrated assessment models (IAMs) such as GLOBIOM~\cite{Havlik2014}, IMAGE~\cite{Stehfest2014}, and MAgPIE~\cite{LotzeCampen2008} have illuminated the possible consequences of alternative technology, land use, and diet scenarios for GHG emissions, biodiversity, and food security. Yet, their complexity and extensive parameterization can make it difficult to parse out causal mechanisms and feedback loops, or to transparently compare policy levers on a conceptual level.
On the theoretical side, dynamic and system-analytic models have provided key insights into food system tipping points, feedback loops between agricultural intensification and land expansion, and the implications of self-reinforcing land degradation~\cite{Bastiaansen2020, coronese2023, Brander1998, scheffer2000}. Land use and food system theories drawing on non-linear dynamics, spatially explicit frameworks, and ecosystem service feedback loops have highlighted the risks of unintended consequences and compensatory effects across spatial and management scales~\cite{Meyfroidt2018, fahrig2003, bengochea2022}. Notably, archetype-based and stylized models have clarified how yield gains can sometimes induce additional land clearing (“rebound” or Jevons effects) rather than sparing nature, depending on feedback loops among value chains, governance, and behavioral factors~\cite{Angelsen2010, Lambin2013, henderson2021unequal}. Complementing this are explorations of “land sparing” versus “land sharing” paradigms, which have underlined the limitations of both simplistic intensification and extensive conservation solutions in isolation~\cite{Phalan2011, Fischer2014}.

Building on this theoretical tradition, our study makes several distinctive contributions to the debate. We present a transparent, parsimonious  model that connects exogenous food demand trajectories to crop and livestock production, land conversion, and feedback from ecosystem integrity that influences yields and future land uses. Such an approach allows us to systematically explore the interplay of demand growth, management intensity, and policy intervention in shaping landscape outcomes and ecological stability.

Key strengths of our work include a calibration to observed, contemporary global trends—most notably the rapid growth in meat demand occurring alongside sustained improvements in average crop yields. By adopting a global perspective, our analysis controls for indirect land-use change effects~\cite{zu2024sustainable}. Furthermore, our framework enables both a realistic assessment of future baseline trajectories and an exploration of plausible supply- and demand-side alternatives, moving beyond purely theoretical scenarios or country case studies.

Furthermore, our spatially explicit landscape component captures the heterogeneity and emergent properties of coupled socio-ecological dynamics, allowing us to assess how context-specific degradation, and restoration processes aggregate into world-scale indicators. This integrated framework is particularly well suited to investigating “policy mixes” and their emergent synergies or trade-offs—issues that remain underexplored yet are highly salient for real-world sustainability transitions~\cite{Rockstrom2020, Clark2020}.

Within this framework, we directly address whether supply-side productivity gains  can meet future food demand while protecting global ecosystems, or if coordinated policy interventions that also target consumption behaviors are essential to achieving long-term sustainability. By illuminating the dynamic, cross-scale feedbacks underpinning both successes and failures in historical and prospective management strategies, our findings contribute to a growing, more theoretically grounded understanding of the pathways—and pitfalls—toward a sustainable food future. 

\begin{figure*}[t!]
\centering
\includegraphics[width=\textwidth]{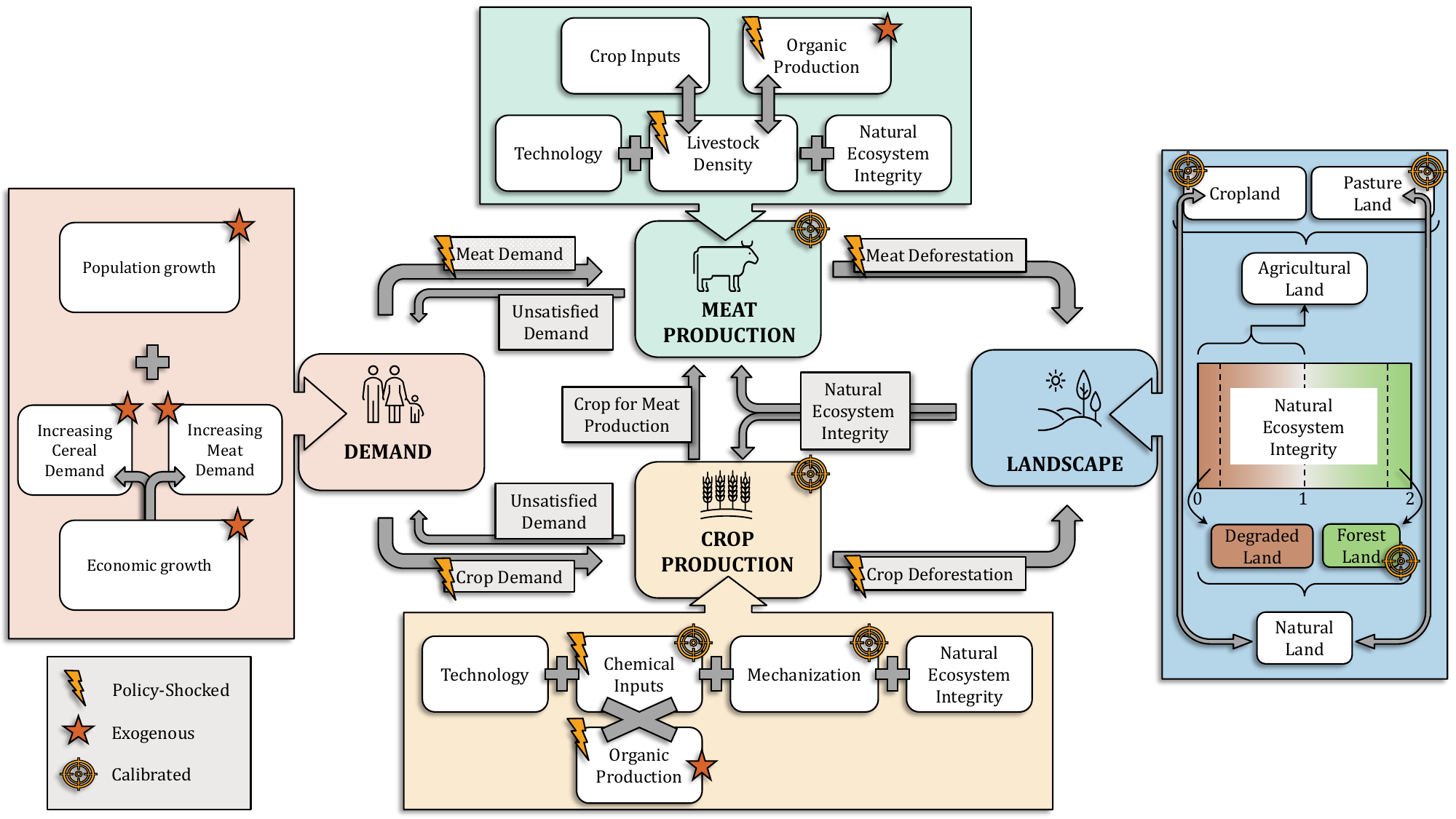}
\caption{Schematic overview of the our dynamic simulation model organized into four main sections: the left panel introduces the demand drivers; the top and bottom panels detail the production and ecological feedback processes; and the right panel summarizes the resulting landscape structure and transitions. Specifically, the left panel shows how external drivers—specifically, population and economic growth (marked with stars for exogenous variables)—jointly determine increases in both cereal and meat demand. Central panels illustrate how these demand trajectories feed into crop and meat production systems, which are influenced by intensification variables such as technology advancement, chemical inputs, mechanization, livestock density, and the adoption of organic production (with targets indicating calibrated variables and lightning bolts indicating policy-shocked variables).
Meat and crop production outcomes depend on and feedback to the integrity of natural ecosystems (a synthetic measure of local biodiversity and soil health, see Section~\ref{sec:ecosystem}), with low integrity reducing yields and increasing future land needs. The model tracks unsatisfied demand for both crops and meat, which triggers further land conversion—expanding either cropland or pasture at the expense of natural land and semi-natural vegetation. The right panel details landscape structure and transitions among key land categories: cropland, pastureland, natural land, forest, and semi-natural vegetation. 
Feedback loops connect ecosystem integrity not only to landscape structure, but also to production system performance, thus enabling the model to capture cross-scale dynamics and the effects of multiple policy levers. 
}
\label{fig:scheme}
\end{figure*}

\section{Model}

\subsection{Model Overview}

We have developed a spatially explicit, discrete-time model that dynamically links the growth in food demand to agricultural production, land conversion among primary land-cover types — namely cropland, pasture, and natural land — and crucially incorporates feedback loops from natural ecosystem integrity (Fig.~\ref{fig:scheme}). The model landscape is represented as a square grid where each cell's land-cover class changes over time, allowing us to track both spatial configuration and aggregate stocks of land-use types. This enables robust quantification of processes such as deforestation, natural regeneration, and degradation pressure as they unfold over time.

The drivers of food demand are exogenous time series that combine projected population growth, changes in per-capita income, and a parameter reflecting dietary preference for meat. This framework is flexible enough to accommodate demand-side policy interventions, such as shifts to lower-meat diets, by perturbing these demand trajectories. 

Agricultural production in our model is governed by both land area and the intensity of management. Cropland yields respond to chemical intensification and mechanization while livestock output depends on available pasture area and on an adjustable livestock-density variable. If, at any time step, aggregate production fails to meet demand, the system responds first by intensifying production (for example, by increasing mechanization and chemical inputs, or livestock density). Persistent shortfalls inevitably trigger the conversion of natural land to cropland or pasture, resulting in deforestation and loss of critical habitat. 

Central to the model is the ecosystem integrity index, a composite indicator reflecting both soil health and biodiversity. This index decreases with intensified mechanization and chemical use but  recovers through periods of natural restoration. The integrity trajectories follow exponential decay and recovery processes, each with a characteristic time scale. Importantly, ecosystem integrity feeds back into agricultural yields, while overexploitation undermines future productivity, thereby creating self-limiting dynamics. In addition, the remaining intact vegetation modulates the pace of ecosystem degradation and restoration at the landscape level through the provision of vital ecosystem services, such as water filtration, climate stability, and biodiversity storage.

 The model incorporates a sustainable alternative to conventional farming through the inclusion of organic management. The choice to focus on organic systems is motivated by their clear institutional definition and the availability of consistent global datasets on organically managed land~\cite{rigby2001organic, michelsen2001recent}. In the model, the transition to organic farming is treated as an exogenous process—that is, the share of organic land is calibrated to empirical data and provided as an input rather than emerging endogenously from the system’s dynamics. Organic management is characterized by the absence of synthetic chemical inputs and stricter limits on livestock density, thereby reducing pressure on ecosystems, albeit often at the cost of short-term yield reductions. The model explores the implications of policy-driven increases in organic adoption and broader reductions in chemical input use, assessing their combined effects on production and ecosystem integrity.

Our framework is capable of contrasting a business-as-usual development trajectory with a suite of alternative policy scenarios, where changes are implemented as constraints or shifts to key model parameters. These include limiting livestock density, restricting deforestation, reducing chemical inputs, and implementing demand-side interventions. This architecture allows us to systematically assess the synergies and trade-offs inherent in: (i) supply-side sustainability measures, (ii) combinations of supply- and demand-side interventions, and (iii) the overall scale of consumption habit transformation needed to achieve a transition onto ecologically sustainable pathways for global land use and food production.

\subsection{Demand: Economic and Population Growth}

Population ($N_t$) and per-capita income ($I_t$) are specified as exogenous drivers that change throught time $t$. Income determines both the total caloric requirement and the composition of human diets. Total caloric demand is modeled using a logarithmic (Engel-type) relationship~\cite{chai2010retrospectives} :

\begin{equation}
\label{eq:calories_demand}
D_t = (a + b\log I_t) N_t ,
\end{equation}
where $D_t$ is total caloric demand, and $a$ and $b$ are positive parameters. The logarithmic form is widely used in empirical studies of food demand because it captures the diminishing marginal increase of total caloric intake with rising income \citep{banks1997quadratic}.

Demand for animal products is modeled as a power-law function of income:
\begin{equation}
\label{eq:meat_demand}
  D_{\mathrm{m}, t} = c\, I_t^{\,d}\,N_t ,
\end{equation}
with $c,d>0$, where $d$ represents the income elasticity of meat demand. Power-law (or constant-elasticity) functions are commonly adopted in agricultural economics to represent the empirically observed sublinear increase in meat and high-value food consumption with income\cite{chang1977functional, gallet2010income}.

The demand for crops directly consumed by humans is the residual once the caloric contribution of meat is subtracted:
\begin{equation}
  D^{\mathrm{food}}_{\mathrm{c}, t} = D_t - D_{\mathrm{m}, t}.
\end{equation}
Then, total crop demand in the system comprises both food and feed uses:
\begin{equation}
  D_{\mathrm{c}, t} = D^{\mathrm{food}}_{\mathrm{c}, t} + D^{\text{feed}}_{\mathrm{c}, t},
\end{equation}
where $D^{\text{feed}}_{\mathrm{c}, t}$ is the requirement of the livestock sector for feed inputs (see Section \ref{sec:production}).
To mimic real-world adaptive demand responses, the model incorporates a feedback mechanism in which consumption is partially suppressed if prior output falls short of demand—analogous to price feedback in markets:
\begin{equation}
  D_{\mathrm{m}, t} \leftarrow \frac{D_{\mathrm{m}, t}}{\omega_{\mathrm{m}}}\left[1 - \alpha_\mathrm{m}\,
    \frac{D_{\mathrm{m}, t-1}-Q_{\mathrm{m}, t-1}}{D_{\mathrm{m}, t-1}}\right],
\end{equation}
\begin{equation}
  D^{\mathrm{food}}_{\mathrm{c}, t} \leftarrow \frac{D^{\mathrm{food}}_{\mathrm{c}, t}}{\omega_{\mathrm{c}}}\left[1 - \alpha_\mathrm{c}\,
    \frac{D_{\mathrm{c}, t-1}-Q_{\mathrm{c}, t-1}}{D_{\mathrm{c}, t-1}}\right],
\end{equation}
where $Q_{\mathrm{m}, t-1}$ and $Q_{\mathrm{c}, t-1}$ denote last period's meat and crop output, respectively, and $\omega_\mathrm{m}$, $\omega^{\mathrm{food}}_{\mathrm{c}}$ are normalization constants. The notation $\leftarrow$ indicates that the exogenously determined demand at time $t$ is endogenously adjusted in response to supply shortfalls in $t-1$. Lastly, normalizing by $\omega_\mathrm{m}$ and $\omega_{\mathrm{c}}$ ensures appropriate scaling of global demand within our stylized representation of the Earth system, while preserving the overall trend -- see Appendix~\ref{app:landscape_init} for further details.

\subsection{Production: Crop and Livestock Systems} \label{sec:production}

Our model incorporates both conventional and organic production systems, which together determine the total outputs of crops and livestock at each time step:
\begin{align}
\label{eq:demand}
Q_{\mathrm{c}, t} = q_{\mathrm{c}, t} + q^{\mathrm{o}}_{c, t} \qquad
Q_{\mathrm{m}, t} = q_{\mathrm{m}, t} + q^{\mathrm{o}}_{m, t},
\end{align}
where $q_{\cdot,t}$ and $q^{\mathrm{o}}_{\cdot, t}$ denote conventional and organic outputs respectively. The share of land under organic management increases exogenously, while both systems are managed by adaptive rules simulating a representative farmer who adjusts input intensities in response to changing economic and environmental conditions.

For conventional crop production, total yield ($q_{\mathrm{c}, t}$) is modeled as a multiplicative function of cultivated area ($A_{\mathrm{c}, t}$), technological level ($T_t$), chemical inputs ($\Phi_{\mathrm{c}, t} \geq 1$), mechanization ($M_{\mathrm{c}, t}\geq 1$), and the condition of the natural ecosystem ($0<\bar{\varepsilon}_{\mathrm{c}, t}\leq 1$):
\begin{equation}
\label{eq:prod_func}
q_{\mathrm{c}, t} = A_{\mathrm{c}, t} (1+T_{t}) (\Phi_{\mathrm{c}, t})^{k} (M_{\mathrm{c}, t})^{f} (\bar{\varepsilon}_{\mathrm{c}, t})^{1-k-f}
\end{equation}
The exponents $k < 1$ and $f < 1$ determine the sensitivities of yield to chemical and mechanical intensification, while $\bar{\varepsilon}_{\mathrm{c}, t}$ captures the impact of average ecosystem integrity on conventional cropland. 

This multiplicative function follows the long tradition of agricultural production modeling (e.g., Cobb–Douglas and Mitscherlich-type functions, see \cite{Mitcherliche_prod_func}), which assume proportional rather than additive contributions of inputs. Including chemical inputs and mechanization reflects historical evidence that these were the dominant drivers of yield growth during the Green Revolution \citep{Evenson2003, Pingali2012}. In contrast, $T_t$ is modeled as an efficiency shifter that increases the productivity of inputs without proportionally raising environmental degradation—for instance, through innovations in precision agriculture or improved crop varieties (see Eq.~\ref{eq:deg_eco_integrity}). This follows the common interpretation of technology as a residual factor capturing yield gains not directly attributable to input intensification~\cite{pierce1999, soto2019}.

Total organic crop yield, which forgo chemical inputs, is given by:
\begin{equation}
\label{eq:prod_func_org}
q^{\mathrm{o}}_{\mathrm{c}, t} = A_{\mathrm{c}, t} (1+T_{t}) (M_{\mathrm{c}, t})^{f} (\bar{\varepsilon}^{\,\mathrm{o}}_{\mathrm{c}, t})^{1-k-f}
\end{equation}
where $\bar{\varepsilon}^{\,\mathrm{o}}_{\mathrm{c}, t}$ is averaged only over organic cropland. Since $\Phi_{\mathrm{c}, t}\geq 1$, organic yields are typically lower than those from conventional management.

Conventional meat production follows a similar structure, with intensification achieved through increased livestock density ($\Lambda_{\mathrm{m,t}}$):
\begin{equation}
q_{\mathrm{m}, t} = A_{\mathrm{m}, t} (1+T_{t}) (\Lambda_{\mathrm{m,t}})^{h} (\bar{\varepsilon}_{\mathrm{m}, t})^{1-h}
\end{equation}
where $A_{\mathrm{m}, t}$ is pasture area, $\Lambda_{\mathrm{m,t}}$ is livestock density, and $h$ modulates sensitivity to ecosystem integrity. Under organic livestock management, livestock densities are restricted, and yields are determined as:
\begin{equation}
q^{\mathrm{o}}_{\mathrm{m}, t} = A_{\mathrm{m}, t} (1+T_{t}) (\min\{\Lambda_{\max}, \Lambda_{\mathrm{m,t}}\})^{h} (\bar{\varepsilon}^{\,\mathrm{o}}_{\mathrm{m}, t})^{1-h}.
\end{equation}
Planned meat output generates a demand for crop-derived feed, which is passed to the crop module as:
\begin{equation}
D^{\text{feed}}_{\mathrm{c}, t} = r \left(q_{\mathrm{m}, t} + q^{\mathrm{o}}_{\mathrm{m}, t}\right)
\end{equation}
where $r$ is the crop-to-meat feed conversion factor. If total crop production is insufficient to meet both food and feed demands, livestock production is scaled back accordingly:
\begin{align}
Q_{\mathrm{m}, t} \leftarrow Q_{\mathrm{m}, t} \min \left(1, \frac{Q_{\mathrm{c}, t}}{D^{\text{food}}_{\mathrm{c}, t} + D^{\text{feed}}_{\mathrm{c}, t}}\right)
\end{align}
This minimization ensures that meat output does not exceed the level achievable with available crop resources after accounting for both direct human consumption and animal feed requirements. The symbol $\leftarrow$ indicates that this constraint applies within the same time period, i.e. livestock are assumed to rely on crop production from the current period. While this represents a simplification—real systems include storage and trade—introducing such mechanisms would substantially increase model complexity.

Input intensities (mechanization, chemical use, livestock density) are updated dynamically for both organic and conventional systems as if managed by a representative farmer who adjusts input intensities to reduce gaps between demand and output in each period. This assumption is consistent with standard  adjustment-cost frameworks in agricultural economics and ecological-economic modeling~\cite{coronese2023, heckbert2010agent}.

Formally, adjustment follows a proportional correction rule:

\begin{equation}
M_{t} = M_{t-1} + \beta M_{t-1} \left( \frac{D_{\mathrm{c}, t} - Q_{\mathrm{c}, t-1}}{D_{\mathrm{c}, t}}\right)
\end{equation}
\begin{equation}
\Phi_{t} = \Phi_{t-1} + \delta \Phi_{t-1} \left( \frac{D_{\mathrm{c}, t} - Q_{\mathrm{c}, t-1}}{D_{\mathrm{c}, t}}\right)
\end{equation}
\begin{equation}
\Lambda_{t} = \Lambda_{t-1} + \gamma \Lambda_{t-1} \left( \frac{D_{\mathrm{m}, t} - Q_{\mathrm{m}, t-1}}{D_{\mathrm{m}, t}}\right)
\end{equation}
where $\beta, \delta$, and $\gamma$ set the speed of adaptive intensification. Technology advances exogenously via logistic growth, aligning with standard practice in stylized food–environment models~\cite{lafuite2018sustainable}:
\begin{equation}
T_{t} = T_{t-1} + \nu T_{t-1}\left(1 - \frac{T_{t-1}}{T_{\max}} \right)
\end{equation}
where $\nu$ and $T_{\max}$ respectively determine the rate and the upper bound of technological advance.

\subsection{Landscape and Natural Ecosystem Integrity}
\label{sec:ecosystem}
\begin{figure*}[t!]
\centering
\includegraphics[width=\textwidth]{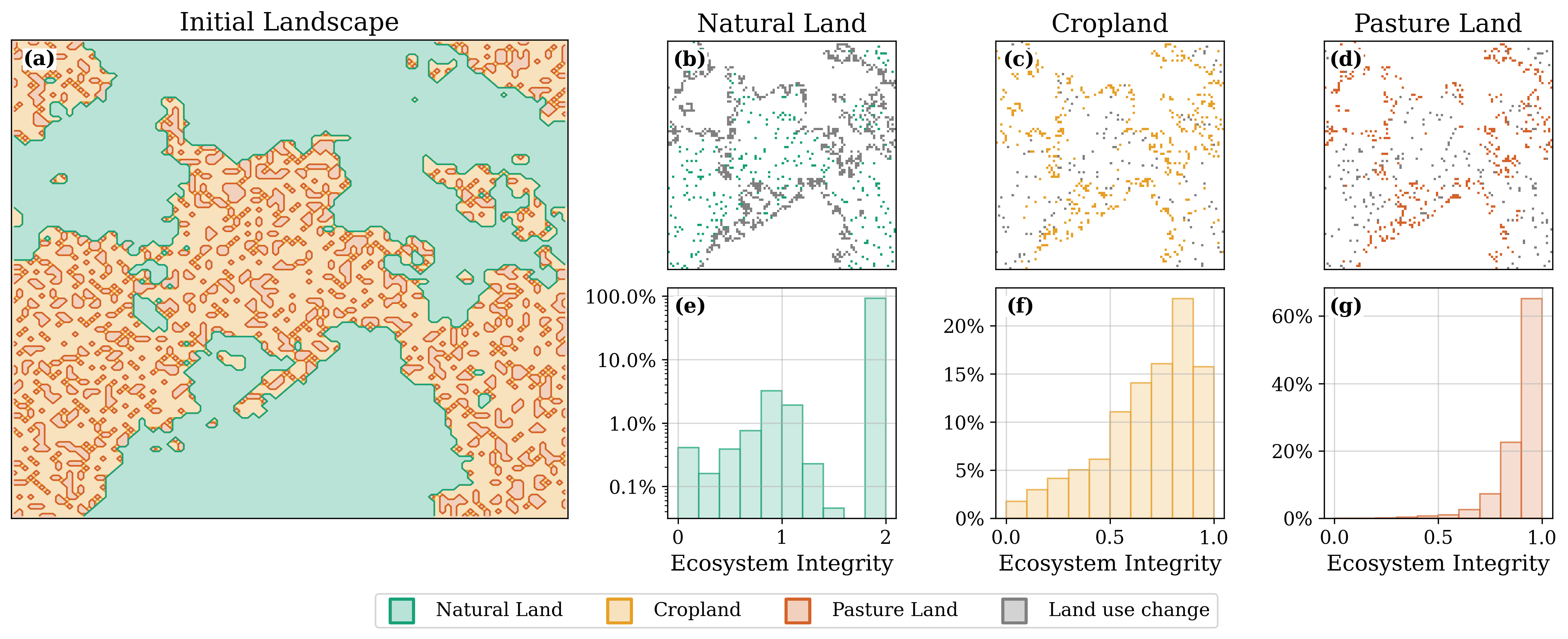}
\caption{Spatial patterns of the initial landscape and subsequent land-use changes. Panel (a) shows the initial arrangement of natural land (green), cropland (yellow), and pastureland (brown), capturing the mosaic structure at the start of the simulation. Panels (b) to (d) illustrate land-use transitions by 2100: green marks new natural land arising from abandonment and restoration, yellow corresponds to recent cropland expansion, brown highlights pastureland conversion, and gray indicates cells that switched from natural land, cropland or pasture land land-use types, respectively,  to another over the simulation period.  As agriculture expands, cropland and pasture increasingly encroach upon forest edges, concentrating in more fertile zones and leaving behind scattered, less productive patches that begin natural restoration. Panels (e) to (g) present histograms of the final distribution of ecosystem integrity for natural land, cropland, and pastureland, respectively. These distributions reveal pronounced degradation in croplands due to continued intensification via mechanization and chemical inputs. Simultaneously, abandoned sites emerge in the natural land distribution, showing early signs of ecological recovery, although these seminatural patches remain well below the ecosystem integrity threshold necessary to qualify as forest ($\varepsilon_{i,t} > 1.9$, see Eq.~\ref{eq:deg_eco_integrity}).}
\label{fig:spatial_explicit}
\end{figure*}

The simulated landscape is represented as a $100 \times 100$ grid\footnote{This system size was chosen to be sufficiently large to smooth out effects of random initialization, while remaining computationally tractable. Results remain consistent when varying grid size, since no variable explicitly depends on system dimensions.}, where each cell is designated as cropland, pasture, or natural land, providing a stylized analogue of real-world land use (Fig.~\ref{fig:spatial_explicit}). The initial allocation reflects global 1960 conditions; over time, land transitions between states as economic and environmental pressures change. 

Natural land may be converted to agriculture if production consistently falls short of demand, whereas underutilized or low-productivity agricultural plots can be abandoned and gradually restored through natural processes. In addition to these demand-driven adjustments, the model incorporates a baseline rate of land expansion and contraction to capture natural turnover in the agricultural sector. This reflects the reality that some farmers exit or enter farming independently of production shortfalls or market conditions, due to personal, institutional, or contextual factors~\cite{katchova2017farm, breustedt2007driving}. Specifically, the rates of expansion ($\varphi^+_{\cdot, t}$) and contraction ($\varphi^-_{\cdot, t}$) of agricultural land are modeled as:

\begin{equation}
    \varphi^+_{\cdot, t} =  \varphi \left\lfloor1 + \zeta^+_{\cdot}\left( \frac{D_{\cdot, t} - Q_{\cdot, t-1}}{D_{\cdot, t}}\right)  \right\rfloor
\end{equation}
\begin{equation}
    \varphi^-_{\cdot, t} =  \varphi \left\lfloor1 + \zeta^-_{\cdot}\left( \frac{Q_{\cdot, t-1} - D_{\cdot, t} }{D_{\cdot, t}}\right)  \right\rfloor
\end{equation}
Here, $\varphi^+{\cdot}$ and $\varphi^-{\cdot}$ represent baseline expansion and contraction rates, while $\zeta^+{\cdot}$ and $\zeta^-{\cdot}$ scale the response to unmet demand or excess production. The notation $\cdot$ is a placeholder for the relevant land type (crop or pasture). 

Sites prioritized for conversion to agriculture are those with highest ecosystem integrity (thus promising higher yields), while marginal, less productive plots tend to be abandoned first, consistent with empirical patterns of agricultural abandonment in more degraded areas~\cite{rey2007abandonment, carreira2024deforestation}.

Each site is assigned a natural ecosystem integrity index ($\varepsilon$), which serves as a synthetic indicator quantifying both above-ground and below-ground ecosystem conditions. Specifically, $\varepsilon$ reflects the multi-dimensional state of ecosystem health, including biodiversity as well as soil health attributes. This approach draws on established ecological frameworks~\cite{de1997multifaceted, watson2018exceptional, blumetto2019ecosystem}, where ecosystem integrity is understood as the capacity of an ecosystem to maintain its characteristic structure, composition, and functional processes in the face of human disturbance. In our model, natural sites have $\varepsilon$ values between 0 and 2, reflecting a gradient from degraded to pristine conditions, whereas agricultural plots are constrained between 0 and 1. Conversion from high-integrity natural to agricultural land causes a sharp drop in $\varepsilon$, representing rapid loss of habitat complexity~\cite{clapcott2012quantifying}. In contrast, restoration or abandonment of agriculture allows for gradual recovery, with sites retaining their existing $\varepsilon$ and accumulating integrity over time. For analytical purposes, we refer to cells with $\varepsilon > 1.9$ as forest and those with $\varepsilon < 0.1$ as degraded—these labels facilitate interpretation but do not alter the underlying dynamics.

Ecosystem integrity in each cell ($\varepsilon_{i,t}$) is updated at every time step using exponential decay, a standard approach in stylized ecological models~\cite{montoya2021habitat, montoya2019trade}. This formulation captures both degradation from agricultural use and natural restoration when land is left fallow. Exponential decay is widely used to represent gradual ecosystem degradation due to sustained pressures, while logistic-style recovery functions mimic the diminishing returns of restoration as ecosystems approach their maximum integrity~\cite{loreau2000biodiversity}. Let $\tau_i$ denote the land-use type of cell $i$—$\mathrm{m}$ for pasture, $\mathrm{c}$ for cropland, and $\mathrm{n}$ for natural land—and let $E_t$ represent the system-wide provision of ecosystem services at time $t$ (see~\ref{eq:ecosystem_service}). The update rule is:

\begin{equation}
\label{eq:deg_eco_integrity}
\varepsilon_{i,t} =  
    \begin{dcases}
    \varepsilon_{i,t-1} \left[ 1 - \theta_{i, \mathrm{m}} \frac{\Lambda_{t}}{{E}_{t}}  \right] & \text{if } \tau_i=\mathrm{m} \\
    \varepsilon_{i,t-1} \left[ 1 - \theta_{i, \mathrm{c}} \frac{M_t + \Phi_t}{{E}_{t}} \right] & \text{if } \tau_i=\mathrm{c} \\
    \varepsilon_{i,t-1} \left[ 1 - \theta_{i, \mathrm{n}} E_t \left(1 - \frac{\varepsilon_{i,t-1}}{\varepsilon_{\max}}\right) \right] & \text{if } \tau_i=\mathrm{n} \\
    \end{dcases}
\end{equation}
Here, $\Lambda_t$, $M_t$, and $\Phi_t$ are the intensities of livestock density, mechanization, and chemical inputs, respectively. The parameters $\theta_{i, \mathrm{m}}$, $\theta_{i, \mathrm{c}}$, and $\theta_{i, \mathrm{n}}$ determine the rates of degradation (for agricultural land) or restoration (for natural land), modulated by overall system intensification and the supply of ecosystem services. $\varepsilon_{\max}$ represents the maximum possible ecosystem integrity.

Ecosystem service supply is captured by a non-linear function of the area and integrity of natural patches in the landscape, drawing on empirical evidence for sublinear (concave) scaling~\cite{loreau2000biodiversity, mitchell2014forest}:
\begin{equation}
\label{eq:ecosystem_service}
    E_t = \left(\frac{\sum_{i\in N} \varepsilon_{i,t}}{\sum_{i\in N} \varepsilon_{i,0}}\right)^p
\end{equation}
where $N$ stands for all natural cells and $p<1$ ensures diminishing returns as the natural area increases.

\subsection{Parameter Specification and Numerical Experiments}

\begin{figure*}[t!]
\centering
\includegraphics[width=\textwidth]{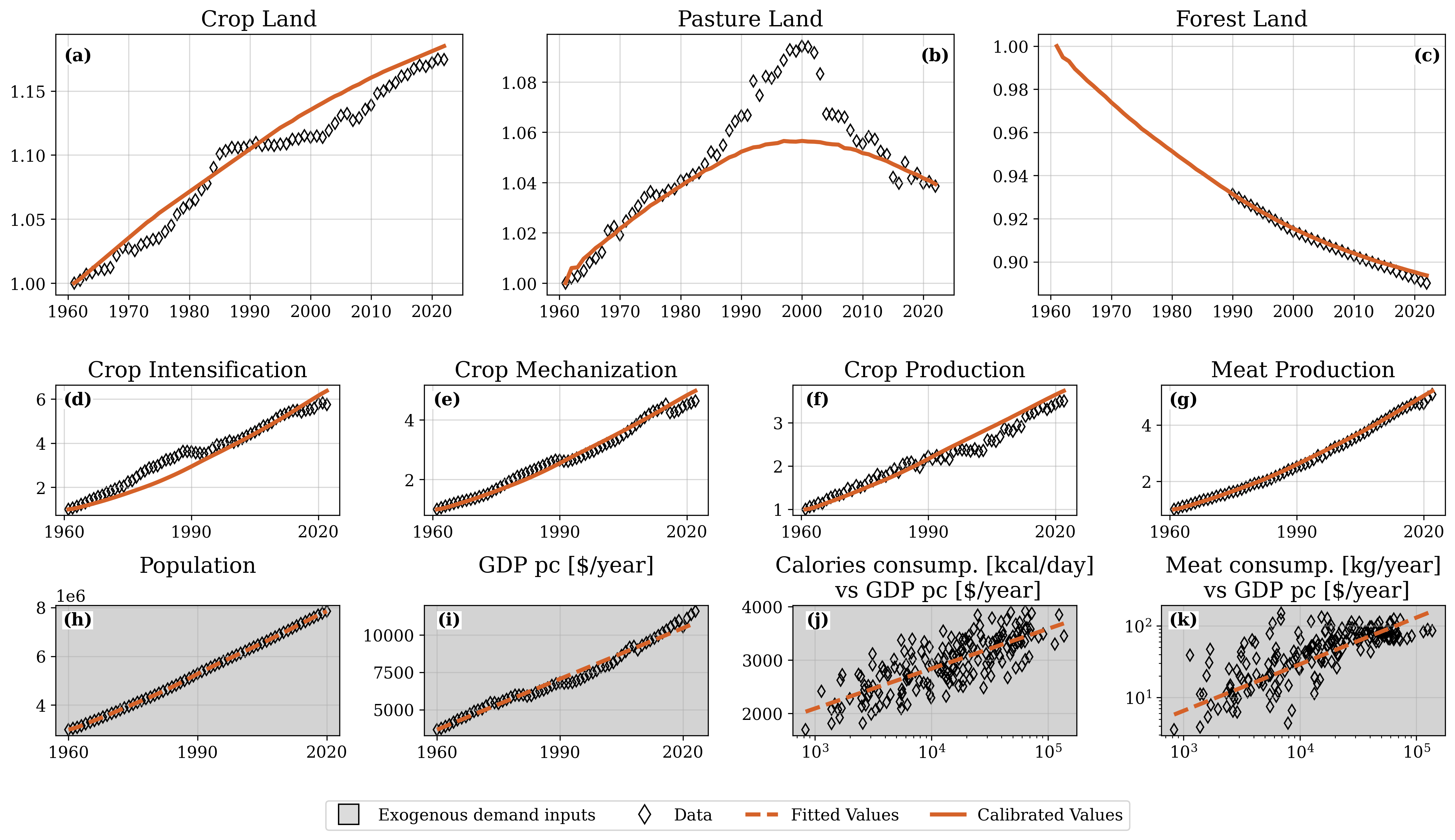}
\caption{Model calibration against major historical trends in agricultural, land use, and socioeconomic indicators.
Each panel displays the model’s simulated results (solid orange lines, normalized to 1 in 1960 to emphasize rates of change) and historical observations from 1960 to 2022 (black diamonds; data from FAO~\cite{faostructuraldata, FAO2024_meat}). The model parameters were calibrated to reproduce empirical trends, and all curves represent averages over 500 Monte Carlo simulations. In particular, the model closely captures key historical patterns, in particular sustained increases in agricultural production and yield, driven by a significant expansion in total demand for both crop and animal products. It also reproduces the peak and subsequent decline in the rate of pasture land expansion, which results from land use efficiency gains as intensification has offset some demand growth. Stable or rising trends in production inputs, such as fertilizer use and mechanization, are also well represented. Gray-shaded panels indicate exogenous input variables (e.g., population, per capita income): here, dashed orange lines represent fitted values matched directly to observed data, which are then used as fixed, absolute model inputs.}
\label{fig:calibration}
\end{figure*}

The model parameters were determined by combining data-driven estimation, calibration against observed agricultural trends, and fixed assignments based on commonly-used values found in the literature (Fig.~\ref{fig:calibration}). Most ecological parameters, including the average degradation and restoration rates ($\theta_{i, \cdot}$), and the ecosystem service exponent ($n$), were primarily set using empirical data or well-established values from previous studies. The same approach was used for the power exponents in the production functions. For these parameters, a more detailed discussion of choice, context specificity, and inherent complexity is provided in the Appendix. 

Of the remaining parameters, those with influence limited to exogenous variables were fitted directly to corresponding empirical data (e.g. demand parameters and the rates of sustainable and organic production adoption). Those that depended on the dynamic realization of the system were calibrated to reproduce observed land use and input trends, particularly using the 1960–2022 FAO data~\cite{faostructuraldata}.

We explored our model through extensive numerical simulations. Unless stated otherwise, results are presented as averages over 500 simulation runs, and standard errors are not shown in the figures as their magnitude was found to be negligible (smaller than  line width). It should be noted that the model does not include stochasticity in the production functions; thus, the observed variation among simulations arises solely from the randomly generated landscape configurations, the stochasticity of restoration and degradation rates, and the random nature of the cells targeted for land-use change.

We evaluated a business-as-usual trajectory alongside a suite of alternative policy scenarios. The baseline scenario was produced by running the model up to the year 2100 with the parameter values listed in Table 1 of the Appendix, while the policy scenarios targeted modifications in the form of constraints or adjusted model parameters. Unless otherwise noted (as in Fig.~\ref{fig:more_sustainable}), all parameter variations in these scenarios represent a 10\% increase or decrease relative to the corresponding calibrated baseline value. For scenarios involving changes in demand, we implemented a 10\% reduction with respect to the additional demand attributed to projected population and income growth between 2022 and 2100. To maintain consistency and comparability between scenarios, every simulation was initialized with all natural ecosystem integrity values set to their maximum value for both natural and agricultural land, and assumed that all land not allocated to agriculture is classified as forest, and thereby excluding the presence of semi-natural vegetation. Although this initial condition may not fully reflect levels of natural degradation in 1960, it is reasonable given that the global expansion of intensive agricultural practices has occurred mainly in recent decades. Moreover, because our primary focus is on outcomes relative to the baseline scenario, this assumption does not substantially affect our overall conclusions. Further information on the initialization procedure can be found in the Appendix.

\section{Results}

The business-as-usual scenario  provides a reference trajectory under current trends in population, consumption habits, and land management practices. This scenario serves as a baseline against which the effectiveness and trade-offs of alternative policy interventions can be assessed.

\begin{figure*}[t!]
\centering
\includegraphics[width=\textwidth]{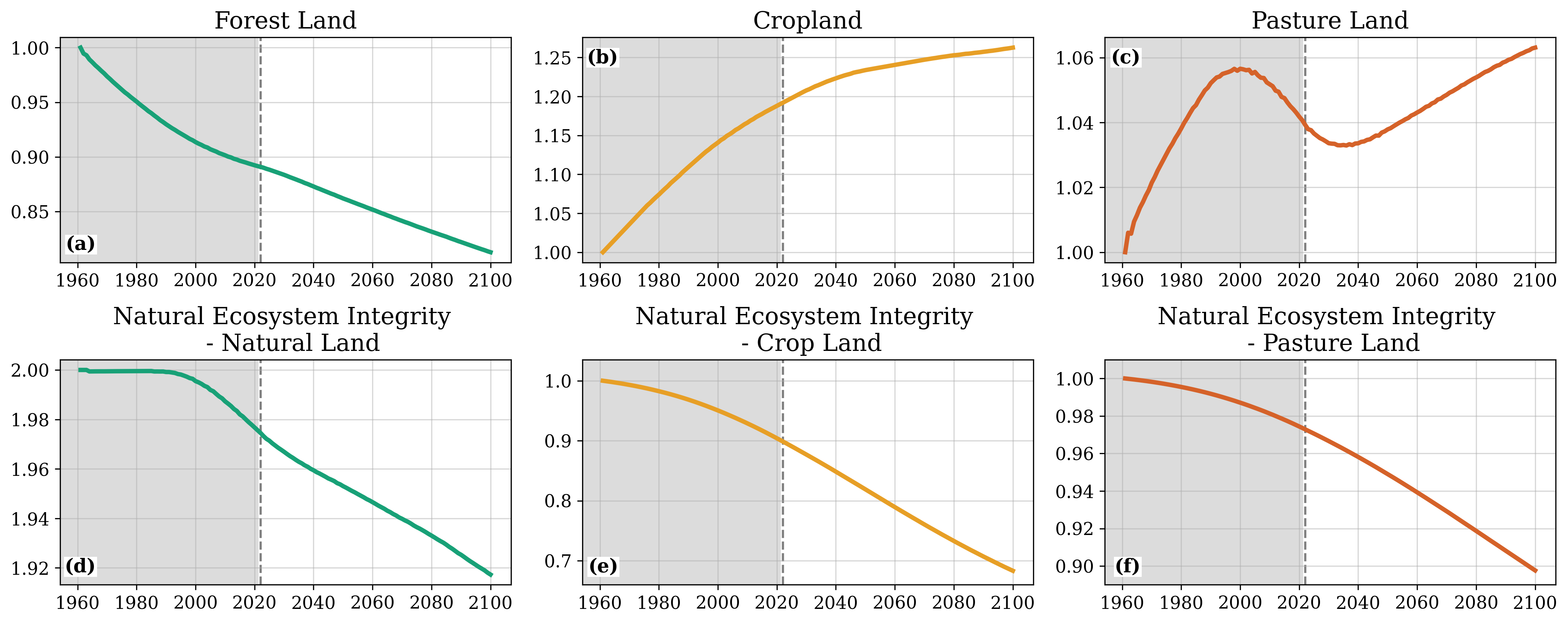}
\caption{Land use and ecosystem integrity trajectories under the business-as-usual scenario. Top panels (a–c) show the temporal trends of the various land-use types: panel (a) tracks the ongoing decline in forest area, panel (b) captures the steady expansion of cropland, and panel (c) traces changes in pastureland. The gray-shaded background indicates the historical period, with projected dynamics shown to the right of the dashed vertical line. Pastureland area initially decreases due to advances in technology and more efficient animal husbandry, which allow for higher livestock densities. However, as technological improvements plateau and can no longer match the rising demand for meat, our model anticipates a renewed expansion of pasture area in the future. Bottom panels (d–f) depict temporal changes in the mean ecosystem integrity index for natural land (d), cropland (e), and pastureland (f). These plots reveal the rapid and sustained degradation of ecosystems associated with agricultural expansion and the ongoing loss of ecological integrity across all land types. Of particular interest is the fact that declines in natural ecosystem integrity lag behind the decrease in forest area; this is because deforestation initially diminishes the supply of ecosystem services vital for maintaining ecosystem stability (see Eq.~\ref{eq:ecosystem_service}). As these services decline, the degradation of agricultural land accelerates (see Eq.~\ref{eq:deg_eco_integrity}), leading to land abandonment and a reduction in the average natural ecosystems integrity.
}
\label{fig:baseline}
\end{figure*}

\subsection{Business-as-usual Scenario}

The business-as-usual scenario highlights a persistent transformation of the landscape, primarily driven by increasing demand for food and livestock products (see Fig.~\ref{fig:baseline}). As agricultural activity intensifies and expands, natural and semi-natural ecosystems are progressively converted or degraded (see Fig.~\ref{fig:baseline}a and Fig.~\ref{fig:baseline}d). Indeed, the expansion of agricultural land directly reduces both the area and overall integrity of natural habitats, resulting in a notable decline in forest land and average ecosystem health within each land use category.

This transformation is particularly evident in the case of cropland (see Fig.~\ref{fig:baseline}b and Fig.~\ref{fig:baseline}e). Here, both land consumption and ecological degradation reach their highest levels, largely as a consequence of the high degree of intensification and mechanization required to meet rising food demands. A substantial proportion ($\approx 40\%$ in recent years) of these crops is allocated to livestock feed, and thus the ecological footprint of crop expansion is strongly tied to meat production and associated changes in dietary habits.

Pastureland dynamics further illustrate the complex feedback mechanisms in the system. While advances in technology and animal husbandry—such as vaccines and growth hormones—have enabled higher livestock densities and contributed to a recent decline in the area under pasture, our model projects that this trend will ultimately reverse. As technological improvements plateau and can no longer keep pace with rapidly rising demand for meat, the pressure for additional pasture expansion resurfaces.

Taken together, these trends accelerate the fragmentation of remaining high-integrity habitats and the spread of degraded patches across the landscape. As demonstrated by spatial analyses (Fig.~\ref{fig:spatial_explicit}), contiguous forest edges are eroded, creating a more fragmented matrix of land uses, while marginal agricultural areas are often abandoned in favor of clearing more productive forested sites. This mosaic of change not only undermines biodiversity but also weakens the ecosystem services that support long-term agricultural productivity.

Collectively, these findings underscore that, without targeted interventions, current trajectories reinforce a self-perpetuating cycle of ecological decline and agricultural expansion. Under this business-as-usual scenario, we observe reduced landscape resilience, increasing dependence on intensification solutions. The baseline scenario therefore highlights the urgent need to explore alternative policy actions.

\subsection{Effects of Single Policy Measures}
We next assess the effects of implementing individual policy interventions, each focused on a specific lever within the food and land system. The complete set of considered scenarios is presented in Fig.~\ref{fig:general_results}, with corresponding parameter details provided in the Appendix. This analysis not only evaluates the direct impact of each measure but also reveals how compensatory system responses can undercut their effectiveness, producing trade-offs or unintended outcomes.

Fig.~\ref{fig:general_results}a maps single-policy scenarios in a space defined by forest land area variation and degraded land variation, both calculated relative to the baseline. These two metrics were selected to capture the dual dimensions of land system change: forest land variation reflects shifts in the extent of intact natural habitats (landscape-scale expansion or protection), while degraded land variation quantifies changes in ecosystem integrity within managed areas due to intensification or restoration efforts. By presenting results in this way, we can compare policies that limit expansion (horizontal direction) with those that primarily aim to improve land already in production (vertical direction), clearly illustrating system-wide trade-offs.

The analysis reveals considerable variability in policy outcomes, much of which stems from compensatory dynamics—where gains in one area trigger pressures elsewhere. For example, supply-side measures such as reducing chemical inputs or expanding organic crop production are effective at lowering land degradation, but typically cause declines in forest area. This is because reduced yields on existing farmland (from chemical restrictions or organic conversion) require compensatory expansion, driving new deforestation and landscape fragmentation to maintain overall production levels. Similarly, implementing deforestation bans for crop or meat production often leads to increased land degradation, because production pressures shift onto existing agricultural land, resulting in more intensive management. Over time, such intensification can diminish yields and degrade ecosystem integrity, which may ultimately drive further land expansion and limit the effectiveness of deforestation policies.

Conversely, demand-focused interventions—particularly reductions in crop or meat consumption—deliver more comprehensive positive outcomes, as they simultaneously relieve pressure for both intensification and expansion. Policies that reduce meat consumption result in the largest gains for forest area and the most pronounced reductions in degraded land. This is because a substantial share of cropland is devoted to growing feed for livestock, and thus plant-based diets reduce both direct and indirect land pressures

\begin{figure*}[t!]
\centering
\includegraphics[width=\textwidth]{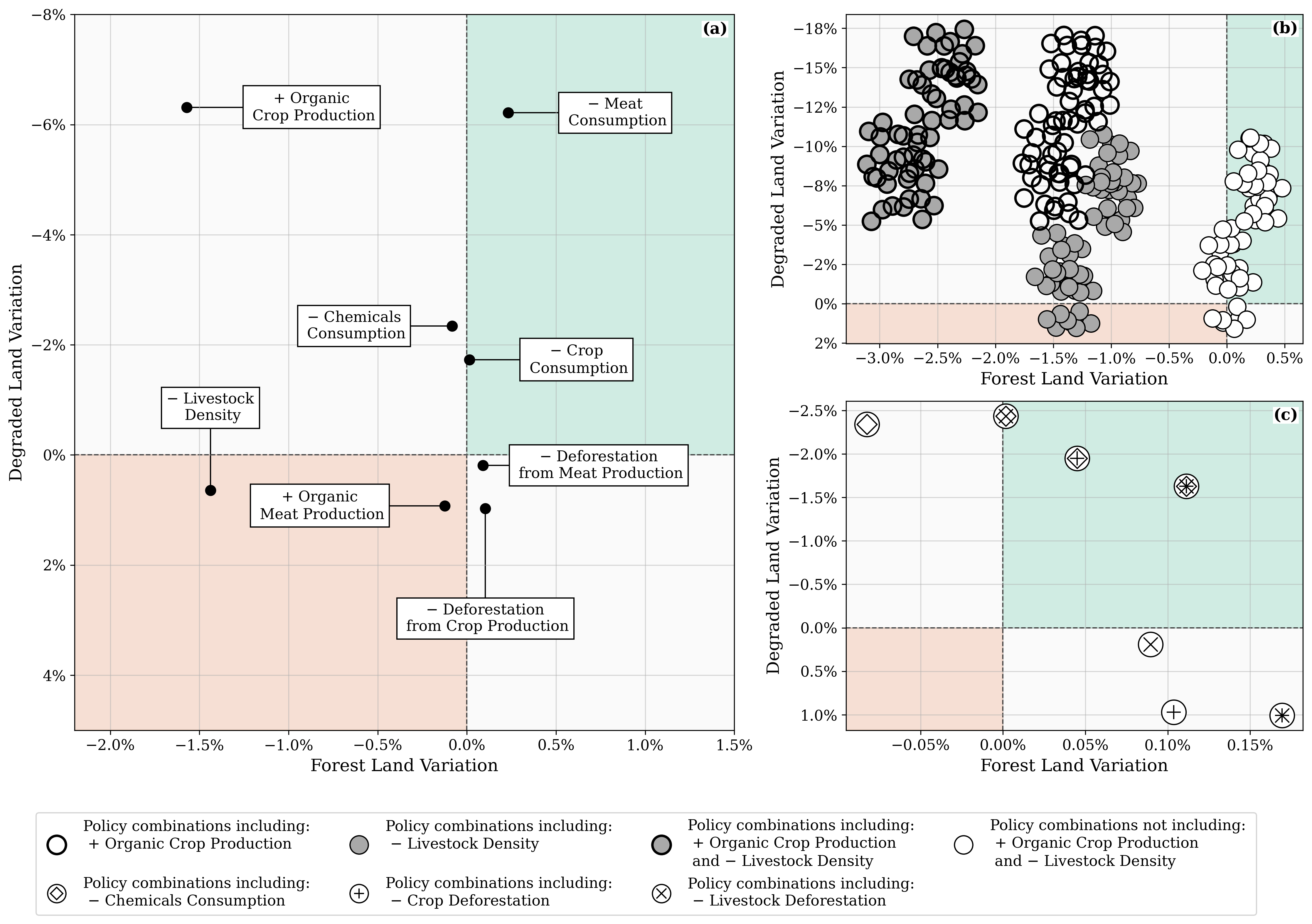}
\caption{Impact of individual and combined policy interventions on forest and degraded area variation. Variations in forest area (horizontal axis) and degraded area (vertical axis) are shown as percentage changes with respect to the baseline scenario. Points in the upper-right green quadrant indicate policy strategies that succeed in both expanding forest area and reducing degraded land, while those in the lower-left quadrant are detrimental on all grounds. Panel (a) shows the effects of individual (single) policy interventions, each marked by a labeled point. Only the reduction in meat consumption achieves positive outcomes on both axes, illustrating that most single interventions are insufficient due to persistent compensatory dynamics within the system. Panel (b) reports outcomes for policy mixes that specifically include increased organic crop production and decreased livestock density—two interventions often viewed as hallmarks of sustainability. Here, each point represents a different mix with other measures. The results demonstrate that these strategies, even when paired with moderate reductions (10\%) in crop and meat demand, consistently fail to halt forest loss, often resulting in further declines in forest land area. Panel (c) highlights selected policy mixes containing key interventions, such as chemical use reduction or restrictions on crop and livestock-driven" deforestation, with distinct symbols denoting which measures are included. This panel shows that policy mixes combining interventions that target both intensification and expansion (i.e. orthogonal strategies) are situated in the green quadrant, signifying that integrated approaches are most effective at overcoming compensatory effects. }
\label{fig:general_results}
\end{figure*}

\section{Synergies and Trade-Offs in Policy Portfolios}

\begin{figure*}[t!]
\centering
\includegraphics[width=\textwidth]{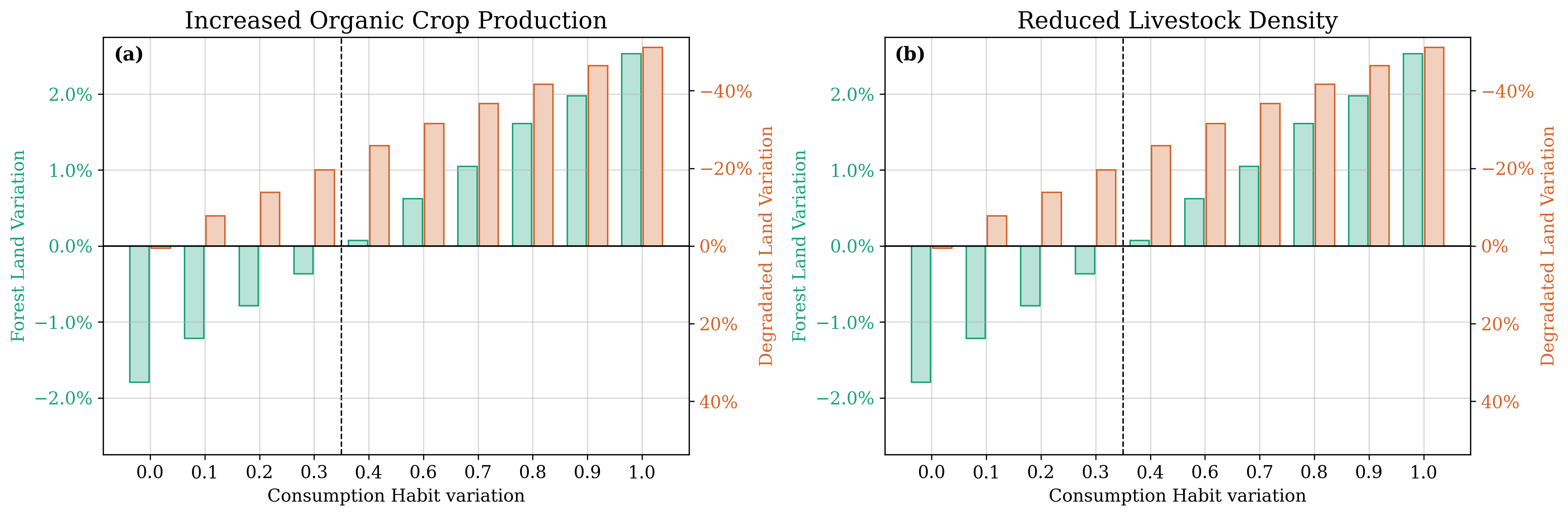}
\caption{Effect of demand reductions on forest and degraded land areas under enhanced organic crop production and reduced livestock density policies. Each panel shows the percentage change in forest land (green, left axis) and degraded land (orange, right axis) compared to the baseline, across scenarios with progressive demand reductions. In particular,  the x-axis represents increasing reductions in projected crop and meat demand by 2100, where  $0$ means no change from business-as-usual and $1$ means no demand increase after 2022.  The results demonstrate that only when demand reductions surpass roughly 40\% do both forest expansion and significant decreases in degraded land occur.}
\label{fig:more_sustainable}
\end{figure*}

\begin{figure*}[t!]
\centering
\includegraphics[width=\textwidth]{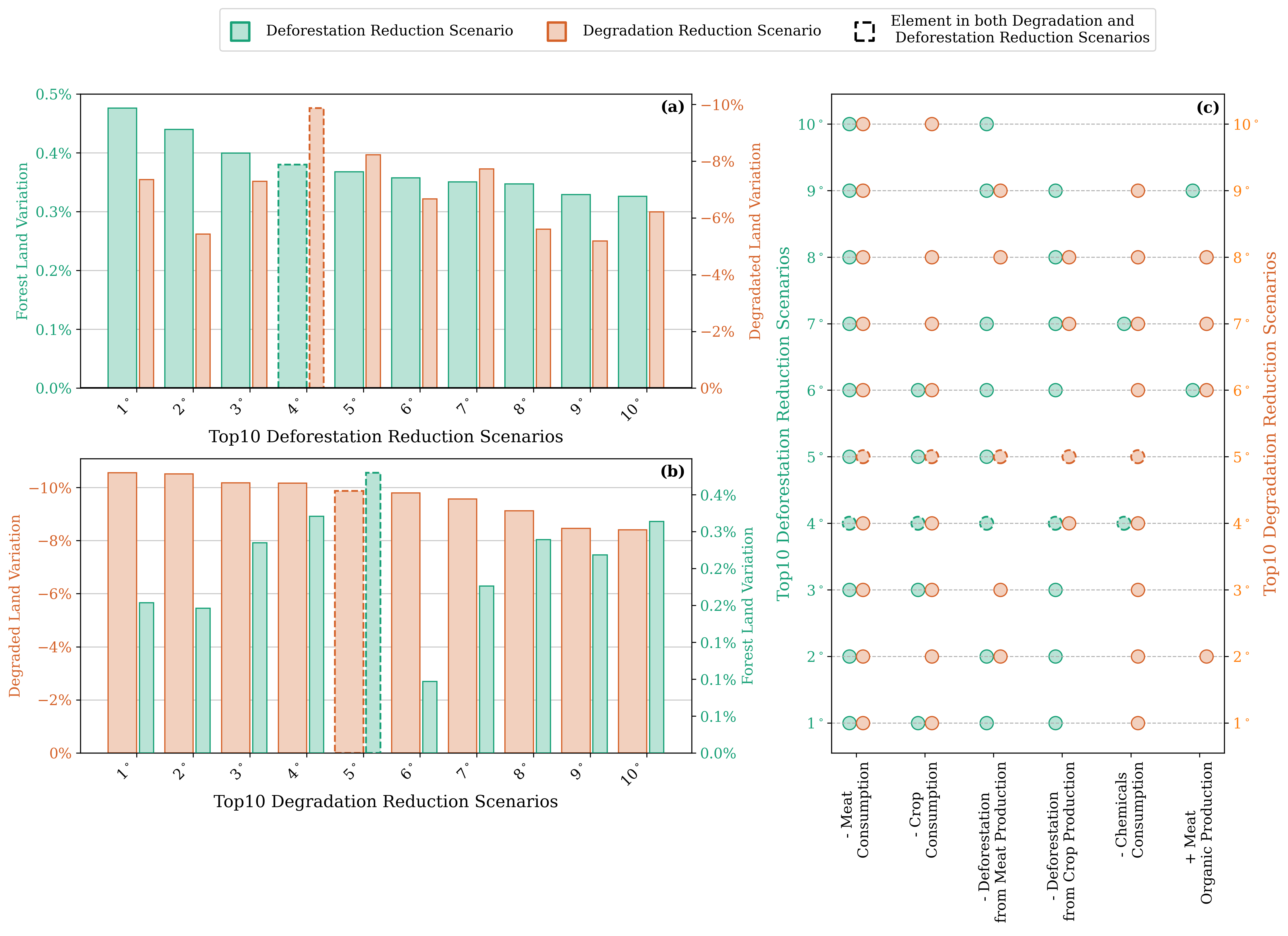}
\caption{Comparative effectiveness of top-performing policy mixes for reducing deforestation and land degradation. Panels (a) and (b)  display, for each of the ten best policy combinations, the relative variation in forest land and degraded land areas as a percentage change from the baseline scenario. Policy mixes are ranked separately by their success in reducing deforestation (a) and degradation (b). Dashed bars indicate policy mixes that appear among the top performers for both objectives together. Panel (c) details the intervention composition of each policy mix. Each row represents a ranked policy mix (from most to least effective; top to bottom), and each circle indicates the inclusion of specific interventions for the corresponding scenario (green for deforestation-focused, orange for degradation-focused).  Despite some overlap, the policy measures most effective at reducing deforestation are not always the same as those best at lowering land degradation. Only a policy portfolio achieves high performance in both dimensions, highlighting the importance—and challenge—of integrated strategies capable of delivering simultaneous gains for forest conservation and reduced agricultural ecosystem degradation.} 
\label{fig:top_scenarios}
\end{figure*}

We then turn to the analysis of policy combinations addressing multiple drivers at once, exploring how different mixes of interventions impact ecosystem integrity. Figures~\ref{fig:general_results}b,~\ref{fig:general_results}c,~\ref{fig:top_scenarios} and~\ref{fig:more_sustainable} summarize these results, allowing direct comparison of policy portfolios.

An initial finding, visible in Figure~\ref{fig:general_results}b, is that policies often regarded as highly sustainable—namely, widespread adoption of organic crop production (eliminating chemical inputs) and reduced livestock intensification (for improved animal welfare)—consistently result in a declining forest land area, regardless of whether they are combined with moderate reductions (10\%) in crop and meat demand.

The underlying mechanisms behind this result differ between crop and meat systems. For organic crop production, the primary constraint is related to the strong diminishing returns ($k=0.2$ in Eq.~\ref{eq:prod_func} and Eq.~\ref{eq:prod_func_org}) from chemical inputs. Moderate chemical use can maintain relatively high yields, but if chemical inputs are eliminated entirely, yield losses grow disproportionately and become very difficult to offset. Most farmers are unable to fully compensate for these losses through additional intensification or mechanization, as the required extra effort would be prohibitive—leading the system to expand cultivation into natural areas to maintain production targets.

The dynamics differ for organic meat production. Here, shifting to organic practices does not generate the same drastic loss of productivity per unit area because industrial animal production is less constrained by input intensification—modern techniques like vertical livestock integration (as seen recently in China~\cite{han2006vertical, wang2023vertical}) support continued output growth. Consequently, organic meat policy mixes exert only modest negative effects on forest land. However, if livestock production is broadly de-intensified without such adaptive capacity, the pressure to maintain meat supply quickly translates into greater expansion of pasture or feed crops, significantly amplifying deforestation.

While some past studies have stopped at these findings~\cite{Fischer2014}—concluding that sustainable production is unfeasible—our model enables quantification of the scale of demand reduction necessary to reverse these negative trends. As shown in Figure~\ref{fig:more_sustainable}, only scenarios with at least a 40\% reduction in projected crop and meat demand by 2100 achieve net gains in forest cover and notable reductions in degraded land when either an increase of organic crop production or a reduction in livestock density is promoted. This highlights the scale of lifestyle and societal change required for promoting animal welfare and eliminating chemical use.

Focusing next on policy mixes that exclude organic crop expansion and livestock intensity reduction, the analysis reveals that combining interventions yield more favorable outcomes than isolated policies. As demonstrated in Fig.~\ref{fig:general_results}c, mixes incorporating at least two different policies outperform single interventions. When policies target both degradation (e.g. reduced chemical use) and deforestation (e.g. conversion controls) together, their effects reinforce one another and mitigate compensatory expansion or intensification. Furthermore, these policies can initiate a positive feedback loop: by halting deforestation, ecosystem service provision is enhanced, and when combined with reduced intensification, this leads to improved natural ecosystem integrity. As ecosystem health recovers, agricultural yields are boosted, which in turn reduces the need for future land expansion.

Examining the highest-performing mixes (Figure~\ref{fig:top_scenarios}), we see that the top portfolios for reducing degradation mainly involve a combination of reduction in chemical inputs alongside decrease in crop and meat consumption—pointing again to the necessity of shifting dietary habits. Conversely, the best-performing portfolios for forest preservation include deforestation restrictions and meat demand reduction, while measures aimed solely at reducing chemicals or crop demand are less central—since they contribute minimally to curbing deforestation, especially that driven by livestock.

Ultimately, only a single portfolio achieves top-tier outcomes for both forest recovery and land degradation, combining deforestation limiting measures, intensification limits, and reductions in both meat and crop consumption. While additional trade-offs may exist beyond the scope of this study, these results highlight that well-designed policy portfolios aligning both supply- and landscape-level actions—with a strong emphasis on demand reduction—are essential for achieving “win-win” gains and real progress toward sustainability.

\section{Discussion}

\subsection{Beyond the Supply-Side: Systemic Barriers to Sustainability}

Our results highlight important limits to achieving a sustainable food system through supply-side reforms alone, echoing findings across the  literature on land use, agricultural intensification, and food system sustainability~\cite{Foley2011,Rockstrom2020,Searchinger2019}. Measures such as reducing chemical inputs and adopting more animal-friendly or organic production systems  provide meaningful improvements at the farm scale—ranging from reductions in local pollution to gains in animal welfare and soil health~\cite{Reganold2016, Springmann2018}. These changes are valuable in themselves, and often come with additional social benefits, including positive impact on human health and responsiveness to consumer demand for more sustainable food options.

However, our model suggests that when per-hectare productivity is deliberately limited—whether by banning chemical inputs or reducing livestock density—the system tends to compensate by converting additional natural or semi-natural land to agriculture. This compensatory expansion can reduce or even offset the environmental benefits achieved on existing farmland, and in some cases may exacerbate pressures such as deforestation, landscape fragmentation, and ecosystem degradation, consistent with previous observations about indirect land-use change~\cite{Searchinger2008, Lambin2013}. Even when multiple supply-side measures are combined, our analysis indicates that their cumulative impacts alone are unlikely to halt forest loss or secure broad-based ecosystem integrity.

Let us stress that these findings do not imply that supply-side reforms are undesirable; rather, they highlight the need to complement them with demand-side transformations and broader systemic changes if the goal is to achieve consistent sustainability.

\subsection{Structural Need for Demand Reductions}

A second, and more profound, implication of our findings is that only strong demand-side interventions are capable of delivering transformative change. Substantial reductions in total food—especially meat—consumption relieve system-wide pressure by cutting both the intensification and expansion drivers simultaneously. This is consistent with a growing body of research indicating that dietary shifts toward reduced animal-source foods (i.e., more plant-based diets) are critical to staying within planetary boundaries and achieving global climate, biodiversity, and public health targets~\cite{Willett2019,Poore2018,Springmann2016}.

Our quantitative analysis provides a concrete threshold: unless at least 40\% of projected excess demand for crops and meat is curbed by 2100, even the most ambitious supply-side or reforestation interventions are unlikely to reverse deforestation or halt land degradation (see also~\cite{Bajzelj2014}). Crucially, demand modification is not only more effective at limiting land use changes but also makes it easier for other sustainability policies (like reduced chemical use or organic expansion) to succeed—a synergy recognized in other modeling exercises and global sustainability assessments~\cite{Clark2020}.

\subsection{Integrated Approaches and Systemic Solutions}

Despite the importance of demand-side measures, our results do not dismiss the value of integrated, multi-faceted policy mixes. Rather, they demonstrate that combinations of carefully designed interventions—addressing both production and consumption—yield the most robust improvements for both forest and degraded land areas variation. Such an integrative approach directly mitigates compensatory effects, wherein policies addressing one part of the system inadvertently shift environmental pressures elsewhere unless implemented holistically~\cite{Lambin2013, Meyfroidt2018, Smith2013}. For instance, coupling landscape-level deforestation controls with intensity reduction measures (e.g., limits on chemical inputs) can prevent rebound effects and increase overall ecosystem integrity~\cite{moretti2024mitigating, Campbell2017}.

It is worth noting, however, that many popular narratives around sustainable intensification or organic conversion tend to overemphasize the benefits of any single intervention, without sufficient attention to the cross-scale feedbacks and compensatory land dynamics illuminated by both our results and other studies~\cite{Garnett2013,Perfecto2016}. A successful transition to a food system that maintains ecosystem integrity and food security together will likely require not just technical or agronomic innovation, but deep societal, economic, and behavioral changes~\cite{Rockstrom2020}.

\subsection{Limitations and Unaccounted Aspects}

While we believe our study advances the understanding of systemic trade-offs and leverage points in the pursuit of sustainable food and land systems, it is important to acknowledge several limitations that may influence the interpretation and applicability of our findings—both in terms of modeling assumptions and the practicalities of real-world policy implementation.

Our framework adopts a stylized, spatially explicit global representation. This abstraction allows us to distill essential system dynamics and identify general insights on policy interactions, while avoiding the computational and data demands of fully resolving geographic, ecological, and socio-economic heterogeneity. But, local variations—shaped by climate variability, land tenure, market integration, and governance~\cite{Lambin2013,Meyfroidt2018, Verburg2013}—are not explicitly represented. However, these features could be incorporated in future extensions through nested multi-scale models or spatially explicit downscaling approaches, enabling more context-sensitive analyses~\cite{lippe2019using}.

Similarly, our land conversion and abandonment rules respond primarily to aggregate supply-demand imbalances. This simplification abstracts from mediating effects such as access to infrastructure, transaction costs, or governance quality. While this allows transparent exploration of system-wide drivers, future model iterations could incorporate heterogeneous decision-making rules or probabilistic entry/exit dynamics to capture local socio-institutional variation~\cite{Garnett2013}.

Crop and livestock production modules rely on generic functional forms calibrated to global data. While sufficient for identifying broad trends, these formulations omit local adaptation, site-specific constraints (e.g., soil degradation, pests, water stress), and stochastic variability in yields. Likewise, our management systems currently exclude agroecological or diversified strategies, focusing instead on conventional and organic monocultural systems. Future work could integrate heterogeneity in production technologies, mixed-cropping systems, or resilience-enhancing practices to more realistically capture the spectrum of sustainable management options~\cite{Kremen2012}.

The socio-political dimension of policy implementation is currently represented exogenously, assuming instantaneous and uniform enactment. While this simplification enables clear analysis of policy potential, it neglects real-world constraints such as political resistance, institutional capacity, stakeholder engagement, and uneven enforcement~\cite{Meyfroidt2022,moretti2025farm}. Incorporating adaptive policy implementation, delays, and compliance heterogeneity would allow more realistic assessment of feasible outcomes and the robustness of interventions.

Our model does not differentiate actors in the food system, and therefore omits social and equity dimensions. While this enables focus on aggregate system-level trade-offs, it excludes potential distributional consequences such as food insecurity, livelihood impacts, or shifts in nutrition. Future extensions could integrate agent-based or household-level modules to explore these outcomes under alternative policy scenarios~\cite{vanDijk2020,HLPE2020}.

Climate-related risks and long-term ecological processes are simplified. Feedbacks from extreme events, biodiversity loss, tipping points, and delayed ecosystem recovery are not dynamically represented. This choice was made to retain tractability for global simulations while highlighting general mechanisms. Future work could include stochastic climate shocks, ecological inertia, and recovery lags to capture more realistic ecological dynamics~\cite{Scheffer2001,IPCC2019,Cramer2008}.

Behavioral and demographic dynamics are also exogenous. Food demand does not adapt through social learning, policy feedback, or price-mediated behavior, and population changes and migration are not modeled. These processes are known to shape long-term land-use trajectories~\cite{henderson2023model,bengochea2022}. Incorporating endogenous behavioral adaptation and demographic dynamics would allow more robust exploration of feasible pathways and potential rebound effects~\cite{Leach2020,starbird2016investing}.

Finally, as with all global modeling efforts, results are subject to parameter uncertainty, data limitations, and simplifications of real-world complexity. Our stylized approach enables identification of core system-level interactions and policy synergies, but should be interpreted as generating hypotheses rather than prescriptive solutions. Future work should prioritize multi-scale, participatory, and transdisciplinary models that incorporate climate risk, social equity, behavioral feedbacks, and realistic policy processes, providing the basis for actionable, locally calibrated insights~\cite{Leach2020,Obersteiner2016}.

\section*{Conclusion}

The results of this study highlight the urgent need for a paradigm shift in how we address food system sustainability. Technological and agronomic improvements, while necessary, are insufficient in isolation to meet the challenges posed by rising global food demand and intensifying environmental pressures. Our systemic approach reveals that only integrated strategies—combining ambitious dietary changes with reformed agricultural practices and strengthened landscape conservation—can reconcile the competing demands of productivity, ecosystem integrity, and social equity.

Our research supports the growing consensus in the literature that demand-side interventions must be placed at the center of sustainability and climate agendas~\cite{Willett2019,Godfray2018}. Avoiding counterproductive trade-offs requires a holistic, system-based approach that aligns policies across supply, demand, and governance. At the same time, future progress depends on robust efforts to tailor and implement these solutions to diverse regional, cultural, and socio-economic contexts, as well as on building public support for transformative change.

Going forward, further refinement of integrated models like ours is needed. In particular, future research should focus on incorporating more explicit representations of climate impact, equity dimensions, market and policy dynamics, and the complex processes underlying behavior and dietary shifts. Advancing multi-scale, participatory, and interdisciplinary modeling will be crucial to develop actionable roadmaps for policymakers and stakeholders.

\section*{Acknowledgments}

 This research was conducted within the Econophysics \& Complex Systems Research Chair, under the aegis of the Fondation du Risque, the Fondation de l’Ecole polytechnique, the Ecole polytechnique and Capital Fund Management. Michel Loreau was supported by the TULIP Laboratory of Excellence (ANR-10-LABX-41).

\bibliographystyle{plainnat}
\bibliography{bibl}

\begin{thebibliography}{88}
\providecommand{\natexlab}[1]{#1}
\providecommand{\url}[1]{\texttt{#1}}
\expandafter\ifx\csname urlstyle\endcsname\relax
  \providecommand{\doi}[1]{doi: #1}\else
  \providecommand{\doi}{doi: \begingroup \urlstyle{rm}\Url}\fi

\bibitem[Angelsen(2010)]{Angelsen2010}
Arild Angelsen.
\newblock Policies for reduced deforestation and their impact on agricultural production.
\newblock \emph{Proceedings of the national Academy of Sciences}, 107\penalty0 (46):\penalty0 19639--19644, 2010.

\bibitem[Baj{\v{z}}elj et~al.(2014)Baj{\v{z}}elj, Richards, Allwood, Smith, Dennis, Curmi, and Gilligan]{Bajzelj2014}
Bojana Baj{\v{z}}elj, Keith~S Richards, Julian~M Allwood, Pete Smith, John~S Dennis, Elizabeth Curmi, and Christopher~A Gilligan.
\newblock Importance of food-demand management for climate mitigation.
\newblock \emph{Nature Climate Change}, 4\penalty0 (10):\penalty0 924--929, 2014.

\bibitem[Banks et~al.(1997)Banks, Blundell, and Lewbel]{banks1997quadratic}
James Banks, Richard Blundell, and Arthur Lewbel.
\newblock Quadratic engel curves and consumer demand.
\newblock \emph{Review of Economics and statistics}, 79\penalty0 (4):\penalty0 527--539, 1997.

\bibitem[Bengochea~Paz et~al.(2022)Bengochea~Paz, Henderson, and Loreau]{bengochea2022}
Diego Bengochea~Paz, Kirsten Henderson, and Michel Loreau.
\newblock Habitat percolation transition undermines sustainability in social-ecological agricultural systems.
\newblock \emph{Ecology Letters}, 25\penalty0 (1):\penalty0 163--176, 2022.

\bibitem[Blumetto et~al.(2019)Blumetto, Castagna, Cardozo, Garc{\'\i}a, Tiscornia, Ruggia, Scarlato, Albicette, Aguerre, and Albin]{blumetto2019ecosystem}
Oscar Blumetto, Andr{\'e}s Castagna, Ger{\'o}nimo Cardozo, Felipe Garc{\'\i}a, Guadalupe Tiscornia, Andrea Ruggia, Santiago Scarlato, Mar{\'\i}a~Marta Albicette, Ver{\'o}nica Aguerre, and Alfredo Albin.
\newblock Ecosystem integrity index, an innovative environmental evaluation tool for agricultural production systems.
\newblock \emph{Ecological indicators}, 101:\penalty0 725--733, 2019.

\bibitem[Bone et~al.(2011)Bone, Dragicevic, and White]{Bastiaansen2020}
Christopher Bone, Suzana Dragicevic, and Roger White.
\newblock Modeling-in-the-middle: bridging the gap between agent-based modeling and multi-objective decision-making for land use change.
\newblock \emph{International Journal of Geographical Information Science}, 25\penalty0 (5):\penalty0 717--737, 2011.

\bibitem[Brander and Taylor(1998)]{Brander1998}
James~A Brander and M~Scott Taylor.
\newblock The simple economics of easter island: A ricardo-malthus model of renewable resource use.
\newblock \emph{American economic review}, pages 119--138, 1998.

\bibitem[Breustedt and Glauben(2007)]{breustedt2007driving}
Gunnar Breustedt and Thomas Glauben.
\newblock Driving forces behind exiting from farming in western europe.
\newblock \emph{Journal of Agricultural economics}, 58\penalty0 (1):\penalty0 115--127, 2007.

\bibitem[Campbell et~al.(2017)Campbell, Beare, Bennett, Hall-Spencer, Ingram, Jaramillo, Ortiz, Ramankutty, Sayer, and Shindell]{Campbell2017}
Bruce~M Campbell, Douglas~J Beare, Elena~M Bennett, Jason~M Hall-Spencer, John~SI Ingram, Fernando Jaramillo, Rodomiro Ortiz, Navin Ramankutty, Jeffrey~A Sayer, and Drew Shindell.
\newblock Agriculture production as a major driver of the earth system exceeding planetary boundaries.
\newblock \emph{Ecology and society}, 22\penalty0 (4), 2017.

\bibitem[Carreira et~al.(2024)Carreira, Costa, and Pessoa]{carreira2024deforestation}
Igor Carreira, Francisco Costa, and Joao~Paulo Pessoa.
\newblock The deforestation effects of trade and agricultural productivity in brazil.
\newblock \emph{Journal of development economics}, 167:\penalty0 103217, 2024.

\bibitem[Cassman(1999)]{cassman1999ecological}
Kenneth~G Cassman.
\newblock Ecological intensification of cereal production systems: yield potential, soil quality, and precision agriculture.
\newblock \emph{Proceedings of the National Academy of Sciences}, 96\penalty0 (11):\penalty0 5952--5959, 1999.

\bibitem[Chai and Moneta(2010)]{chai2010retrospectives}
Andreas Chai and Alessio Moneta.
\newblock Retrospectives: engel curves.
\newblock \emph{Journal of Economic Perspectives}, 24\penalty0 (1):\penalty0 225--240, 2010.

\bibitem[Chang(1977)]{chang1977functional}
Hui-shyong Chang.
\newblock Functional forms and the demand for meat in the united states.
\newblock \emph{The Review of Economics and Statistics}, pages 355--359, 1977.

\bibitem[Clapcott et~al.(2012)Clapcott, Collier, Death, Goodwin, Harding, Kelly, Leathwick, and Young]{clapcott2012quantifying}
Joanne~E Clapcott, Kevin~J Collier, Russell~G Death, EO~Goodwin, Jon~S Harding, David Kelly, John~R Leathwick, and Roger~G Young.
\newblock Quantifying relationships between land-use gradients and structural and functional indicators of stream ecological integrity.
\newblock \emph{Freshwater Biology}, 57\penalty0 (1):\penalty0 74--90, 2012.

\bibitem[Clapp(2020)]{HLPE2020}
Jennifer Clapp.
\newblock Food security and nutrition: building a global narrative towards 2030.
\newblock 2020.

\bibitem[Clark et~al.(2020)Clark, Domingo, Colgan, Thakrar, Tilman, Lynch, and et~al.]{Clark2020}
Michael~A. Clark, Norah G.~G. Domingo, Kimberly Colgan, Sam Thakrar, David Tilman, John Lynch, and et~al.
\newblock Global food system emissions could preclude achieving the 1.5 and 2 c climate change targets.
\newblock \emph{Science}, 370\penalty0 (6517):\penalty0 705--708, 2020.

\bibitem[Coronese et~al.(2023)Coronese, Occelli, Lamperti, and Roventini]{coronese2023}
Matteo Coronese, Martina Occelli, Francesco Lamperti, and Andrea Roventini.
\newblock Agrilove: agriculture, land-use and technical change in an evolutionary, agent-based model.
\newblock \emph{Ecological Economics}, 208:\penalty0 107756, 2023.

\bibitem[Cramer et~al.(2008)Cramer, Hobbs, and Standish]{Cramer2008}
Viki~A Cramer, Richard~J Hobbs, and Rachel~J Standish.
\newblock What's new about old fields? land abandonment and ecosystem assembly.
\newblock \emph{Trends in ecology \& evolution}, 23\penalty0 (2):\penalty0 104--112, 2008.

\bibitem[Crouzeilles et~al.(2016)Crouzeilles, Curran, Ferreira, Lindenmayer, Grelle, and Rey~Benayas]{crouzeilles2016global}
Renato Crouzeilles, Michael Curran, Mariana~S Ferreira, David~B Lindenmayer, Carlos~EV Grelle, and Jos{\'e}~M Rey~Benayas.
\newblock A global meta-analysis on the ecological drivers of forest restoration success.
\newblock \emph{Nature communications}, 7\penalty0 (1):\penalty0 11666, 2016.

\bibitem[De~Leo and Levin(1997)]{de1997multifaceted}
Giulio~A De~Leo and Simon Levin.
\newblock The multifaceted aspects of ecosystem integrity.
\newblock \emph{Conservation ecology}, 1\penalty0 (1), 1997.

\bibitem[Evans et~al.(2020)Evans, Quinton, Davies, Zhao, and Govers]{evans2020soil}
DL~Evans, John~N Quinton, Jessica~AC Davies, Jianlin Zhao, and Gerard Govers.
\newblock Soil lifespans and how they can be extended by land use and management change.
\newblock \emph{Environmental Research Letters}, 15\penalty0 (9):\penalty0 0940b2, 2020.

\bibitem[Evenson and Gollin(2003)]{Evenson2003}
Robert~E Evenson and Douglas Gollin.
\newblock Assessing the impact of the green revolution, 1960 to 2000.
\newblock \emph{science}, 300\penalty0 (5620):\penalty0 758--762, 2003.

\bibitem[Evenson and Pingali(2009)]{evenson2009handbook}
Robert~E Evenson and Prabhu Pingali.
\newblock \emph{Handbook of agricultural economics}, volume~4.
\newblock Elsevier, 2009.

\bibitem[Fahrig(2003)]{fahrig2003}
Lenore Fahrig.
\newblock Effects of habitat fragmentation on biodiversity.
\newblock \emph{Annual review of ecology, evolution, and systematics}, 34\penalty0 (1):\penalty0 487--515, 2003.

\bibitem[Fischer et~al.(2014)Fischer, Abson, Butsic, Chappell, Ekroos, Hanspach, Kuemmerle, Smith, and von Wehrden]{Fischer2014}
Joern Fischer, David~J Abson, Van Butsic, M~Jahi Chappell, Johan Ekroos, Jan Hanspach, Tobias Kuemmerle, Henrik~G Smith, and Henrik von Wehrden.
\newblock Land sparing versus land sharing: moving forward.
\newblock \emph{Conservation Letters}, 7\penalty0 (3):\penalty0 149--157, 2014.

\bibitem[Foley et~al.(2011)Foley, Ramankutty, Brauman, Cassidy, Gerber, Johnston, Mueller, O’Connell, Ray, West, et~al.]{Foley2011}
Jonathan~A Foley, Navin Ramankutty, Kate~A Brauman, Emily~S Cassidy, James~S Gerber, Matt Johnston, Nathaniel~D Mueller, Christine O’Connell, Deepak~K Ray, Paul~C West, et~al.
\newblock Solutions for a cultivated planet.
\newblock \emph{Nature}, 478\penalty0 (7369):\penalty0 337--342, 2011.

\bibitem[{Food and Agriculture Organization of the United Nations}(2024)]{FAO2024_meat}
{Food and Agriculture Organization of the United Nations}.
\newblock Yearly per capita supply of all meat [dataset].
\newblock \url{https://archive.ourworldindata.org/20250925-235133/grapher/meat-supply-per-person.html}, 2024.
\newblock Processed by Our World in Data. Original data: "Food Balances: Food Balances (-2013, old methodology and population)" and "Food Balances: Food Balances (2010-)". Retrieved October 1, 2025 (archived on September 25, 2025).

\bibitem[{Food and Agriculture Organization of the United Nations (FAO)}(Accessed 2024)]{faostructuraldata}
{Food and Agriculture Organization of the United Nations (FAO)}.
\newblock Structural data from agricultural censuses.
\newblock FAOSTAT database, Accessed 2024.
\newblock URL \url{https://www.fao.org/faostat/en/#data}.
\newblock Accessed on September 10, 2025.

\bibitem[Gallet(2010)]{gallet2010income}
Craig~A Gallet.
\newblock The income elasticity of meat: a meta-analysis.
\newblock \emph{Australian Journal of Agricultural and Resource Economics}, 54\penalty0 (4):\penalty0 477--490, 2010.

\bibitem[Garc{\'\i}a-Ruiz et~al.(2015)Garc{\'\i}a-Ruiz, Beguer{\'\i}a, Nadal-Romero, Gonz{\'a}lez-Hidalgo, Lana-Renault, and Sanju{\'a}n]{garcia2015meta}
Jos{\'e}~M Garc{\'\i}a-Ruiz, Santiago Beguer{\'\i}a, Estela Nadal-Romero, Jos{\'e}~C Gonz{\'a}lez-Hidalgo, Noem{\'\i} Lana-Renault, and Yasmina Sanju{\'a}n.
\newblock A meta-analysis of soil erosion rates across the world.
\newblock \emph{Geomorphology}, 239:\penalty0 160--173, 2015.

\bibitem[Garnett et~al.(2013)Garnett, Appleby, Balmford, Bateman, Benton, Bloomer, Burlingame, Dawkins, Dolan, Fraser, et~al.]{Garnett2013}
Tara Garnett, Michael~C Appleby, Andrew Balmford, Ian~J Bateman, Tim~G Benton, Phil Bloomer, Barbara Burlingame, Marian Dawkins, Liam Dolan, David Fraser, et~al.
\newblock Sustainable intensification in agriculture: premises and policies.
\newblock \emph{Science}, 341\penalty0 (6141):\penalty0 33--34, 2013.

\bibitem[Godfray et~al.(2018)Godfray, Aveyard, Garnett, Hall, Key, Lorimer, Pierrehumbert, Scarborough, Springmann, and Jebb]{Godfray2018}
H~Charles~J Godfray, Paul Aveyard, Tara Garnett, Jim~W Hall, Timothy~J Key, Jamie Lorimer, Ray~T Pierrehumbert, Peter Scarborough, Marco Springmann, and Susan~A Jebb.
\newblock Meat consumption, health, and the environment.
\newblock \emph{Science}, 361\penalty0 (6399):\penalty0 eaam5324, 2018.

\bibitem[Han et~al.(2006)Han, Trienekens, Tan, Omta, and Wang]{han2006vertical}
J~Han, JH~Trienekens, T~Tan, SWF Omta, and K~Wang.
\newblock Vertical coordination, quality management and firm performance of the pork processing industry in china.
\newblock In \emph{International agrifood chains and networks}, pages 319--332. Wageningen Academic, 2006.

\bibitem[Havl{\'\i}k et~al.(2014)Havl{\'\i}k, Valin, Herrero, Obersteiner, Schmid, Rufino, Mosnier, Thornton, B{\"o}ttcher, Conant, et~al.]{Havlik2014}
Petr Havl{\'\i}k, Hugo Valin, Mario Herrero, Michael Obersteiner, Erwin Schmid, Mariana~C Rufino, Aline Mosnier, Philip~K Thornton, Hannes B{\"o}ttcher, Richard~T Conant, et~al.
\newblock Climate change mitigation through livestock system transitions.
\newblock \emph{Proceedings of the National Academy of Sciences}, 111\penalty0 (10):\penalty0 3709--3714, 2014.

\bibitem[Heckbert et~al.(2010)Heckbert, Baynes, and Reeson]{heckbert2010agent}
Scott Heckbert, Tim Baynes, and Andrew Reeson.
\newblock Agent-based modeling in ecological economics.
\newblock \emph{Annals of the new York Academy of Sciences}, 1185\penalty0 (1):\penalty0 39--53, 2010.

\bibitem[Henderson and Loreau(2021)]{henderson2021unequal}
Kirsten Henderson and Michel Loreau.
\newblock Unequal access to resources undermines global sustainability.
\newblock \emph{Science of The Total Environment}, 763:\penalty0 142981, 2021.

\bibitem[Henderson and Loreau(2023)]{henderson2023model}
Kirsten Henderson and Michel Loreau.
\newblock A model of sustainable development goals: Challenges and opportunities in promoting human well-being and environmental sustainability.
\newblock \emph{Ecological modelling}, 475:\penalty0 110164, 2023.

\bibitem[Katchova and Ahearn(2017)]{katchova2017farm}
Ani~L Katchova and Mary~Clare Ahearn.
\newblock Farm entry and exit from us agriculture.
\newblock \emph{Agricultural Finance Review}, 77\penalty0 (1):\penalty0 50--63, 2017.

\bibitem[Kremen and Miles(2012)]{Kremen2012}
Claire Kremen and Albie Miles.
\newblock Ecosystem services in biologically diversified versus conventional farming systems: benefits, externalities, and trade-offs.
\newblock \emph{Ecology and society}, 17\penalty0 (4), 2012.

\bibitem[Lafuite et~al.(2018)Lafuite, Denise, and Loreau]{lafuite2018sustainable}
A-S Lafuite, Gonzague Denise, and Michel Loreau.
\newblock Sustainable land-use management under biodiversity lag effects.
\newblock \emph{Ecological Economics}, 154:\penalty0 272--281, 2018.

\bibitem[Lambin and Meyfroidt(2010)]{Lambin2013}
Eric~F Lambin and Patrick Meyfroidt.
\newblock Land use transitions: Socio-ecological feedback versus socio-economic change.
\newblock \emph{Land use policy}, 27\penalty0 (2):\penalty0 108--118, 2010.

\bibitem[Leach et~al.(2018)Leach, Reyers, Bai, Brondizio, Cook, D{\'\i}az, Espindola, Scobie, Stafford-Smith, and Subramanian]{Leach2020}
Melissa Leach, Belinda Reyers, Xuemei Bai, Eduardo~S Brondizio, Christina Cook, Sandra D{\'\i}az, Giovana Espindola, Michelle Scobie, Mark Stafford-Smith, and Suneetha~M Subramanian.
\newblock Equity and sustainability in the anthropocene: A social--ecological systems perspective on their intertwined futures.
\newblock \emph{Global Sustainability}, 1:\penalty0 e13, 2018.

\bibitem[Lippe et~al.(2019)Lippe, Bithell, Gotts, Natalini, Barbrook-Johnson, Giupponi, Hallier, Hofstede, Le~Page, Matthews, et~al.]{lippe2019using}
Melvin Lippe, Mike Bithell, Nick Gotts, Davide Natalini, Peter Barbrook-Johnson, Carlo Giupponi, Mareen Hallier, Gert~Jan Hofstede, Christophe Le~Page, Robin~B Matthews, et~al.
\newblock Using agent-based modelling to simulate social-ecological systems across scales.
\newblock \emph{GeoInformatica}, 23\penalty0 (2):\penalty0 269--298, 2019.

\bibitem[Loreau(2000)]{loreau2000biodiversity}
Michel Loreau.
\newblock Biodiversity and ecosystem functioning: recent theoretical advances.
\newblock \emph{Oikos}, 91\penalty0 (1):\penalty0 3--17, 2000.

\bibitem[Loreau(2010)]{loreau2010populations}
Michel Loreau.
\newblock From populations to ecosystems: theoretical foundations for a new ecological synthesis.
\newblock In \emph{From Populations to Ecosystems}. Princeton University Press, 2010.

\bibitem[Lotze-Campen et~al.(2008)Lotze-Campen, M{\"u}ller, Bondeau, Rost, Popp, and Lucht]{LotzeCampen2008}
Hermann Lotze-Campen, Christoph M{\"u}ller, Alberte Bondeau, Stefanie Rost, Alexander Popp, and Wolfgang Lucht.
\newblock Global food demand, productivity growth, and the scarcity of land and water resources: a spatially explicit mathematical programming approach.
\newblock \emph{Agricultural Economics}, 39\penalty0 (3):\penalty0 325--338, 2008.

\bibitem[Mbow et~al.(2017)Mbow, Reisinger, Canadell, and O’Brien]{IPCC2019}
Hans-Otto~P{\"o}rtner Mbow, Andy Reisinger, Josep Canadell, and Phillip O’Brien.
\newblock Special report on climate change, desertification, land degradation, sustainable land management, food security, and greenhouse gas fluxes in terrestrial ecosystems (sr2).
\newblock \emph{Ginevra, IPCC}, 650, 2017.

\bibitem[Meyfroidt et~al.(2022{\natexlab{a}})Meyfroidt, De~Bremond, Ryan, Archer, Aspinall, Chhabra, Camara, Corbera, DeFries, D{\'\i}az, et~al.]{Meyfroidt2018}
Patrick Meyfroidt, Ariane De~Bremond, Casey~M Ryan, Emma Archer, Richard Aspinall, Abha Chhabra, Gilberto Camara, Esteve Corbera, Ruth DeFries, Sandra D{\'\i}az, et~al.
\newblock Ten facts about land systems for sustainability.
\newblock \emph{Proceedings of the National Academy of Sciences}, 119\penalty0 (7):\penalty0 e2109217118, 2022{\natexlab{a}}.

\bibitem[Meyfroidt et~al.(2022{\natexlab{b}})Meyfroidt, De~Bremond, Ryan, Archer, Aspinall, Chhabra, Camara, Corbera, DeFries, D{\'\i}az, et~al.]{Meyfroidt2022}
Patrick Meyfroidt, Ariane De~Bremond, Casey~M Ryan, Emma Archer, Richard Aspinall, Abha Chhabra, Gilberto Camara, Esteve Corbera, Ruth DeFries, Sandra D{\'\i}az, et~al.
\newblock Ten facts about land systems for sustainability.
\newblock \emph{Proceedings of the National Academy of Sciences}, 119\penalty0 (7):\penalty0 e2109217118, 2022{\natexlab{b}}.

\bibitem[Michelsen(2001)]{michelsen2001recent}
Johannes Michelsen.
\newblock Recent development and political acceptance of organic farming in europe.
\newblock \emph{Sociologia ruralis}, 41\penalty0 (1):\penalty0 3--20, 2001.

\bibitem[Milne(1991)]{milne1991utility}
Bruce~T Milne.
\newblock The utility of fractal geometry in landscape design.
\newblock \emph{Landscape and Urban Planning}, 21\penalty0 (1-2):\penalty0 81--90, 1991.

\bibitem[Mitchell et~al.(2014)Mitchell, Bennett, and Gonzalez]{mitchell2014forest}
Matthew~GE Mitchell, Elena~M Bennett, and Andrew Gonzalez.
\newblock Forest fragments modulate the provision of multiple ecosystem services.
\newblock \emph{Journal of Applied Ecology}, 51\penalty0 (4):\penalty0 909--918, 2014.

\bibitem[Montoya et~al.(2019)Montoya, Haegeman, Gaba, De~Mazancourt, Bretagnolle, and Loreau]{montoya2019trade}
Daniel Montoya, Bart Haegeman, Sabrina Gaba, Claire De~Mazancourt, Vincent Bretagnolle, and Michel Loreau.
\newblock Trade-offs in the provisioning and stability of ecosystem services in agroecosystems.
\newblock \emph{Ecological Applications}, 29\penalty0 (2):\penalty0 e01853, 2019.

\bibitem[Montoya et~al.(2021)Montoya, Haegeman, Gaba, De~Mazancourt, and Loreau]{montoya2021habitat}
Daniel Montoya, Bart Haegeman, Sabrina Gaba, Claire De~Mazancourt, and Michel Loreau.
\newblock Habitat fragmentation and food security in crop pollination systems.
\newblock \emph{Journal of Ecology}, 109\penalty0 (8):\penalty0 2991--3006, 2021.

\bibitem[Moretti and Benzaquen(2024)]{moretti2024mitigating}
Elia Moretti and Michael Benzaquen.
\newblock Mitigating farmland biodiversity loss: A bio-economic model of land consolidation and pesticide use.
\newblock \emph{arXiv preprint arXiv:2407.19749}, 2024.

\bibitem[Moretti et~al.(2025)Moretti, Loreau, and Benzaquen]{moretti2025farm}
Elia Moretti, Michel Loreau, and Michael Benzaquen.
\newblock Farm size matters: A spatially explicit ecological-economic framework for biodiversity and pest control.
\newblock \emph{arXiv preprint arXiv:2505.17687}, 2025.

\bibitem[Moss(2022)]{Mitcherliche_prod_func}
Charles~B Moss.
\newblock \emph{Production Economics}.
\newblock World Scientific, 2022.
\newblock \doi{10.1142/12332}.
\newblock URL \url{https://www.worldscientific.com/doi/abs/10.1142/12332}.

\bibitem[Obersteiner et~al.(2016)Obersteiner, Walsh, Frank, Havl{\'\i}k, Cantele, Liu, Palazzo, Herrero, Lu, Mosnier, et~al.]{Obersteiner2016}
Michael Obersteiner, Brian Walsh, Stefan Frank, Petr Havl{\'\i}k, Matthew Cantele, Junguo Liu, Amanda Palazzo, Mario Herrero, Yonglong Lu, Aline Mosnier, et~al.
\newblock Assessing the land resource--food price nexus of the sustainable development goals.
\newblock \emph{Science advances}, 2\penalty0 (9):\penalty0 e1501499, 2016.

\bibitem[Pendrill et~al.(2022)Pendrill, Gardner, Meyfroidt, Persson, Adams, Azevedo, Bastos~Lima, Baumann, Curtis, De~Sy, et~al.]{Pendrill2022}
Florence Pendrill, Toby~A Gardner, Patrick Meyfroidt, U~Martin Persson, Justin Adams, Tasso Azevedo, Mairon~G Bastos~Lima, Matthias Baumann, Philip~G Curtis, Veronique De~Sy, et~al.
\newblock Disentangling the numbers behind agriculture-driven tropical deforestation.
\newblock \emph{Science}, 377\penalty0 (6611):\penalty0 eabm9267, 2022.

\bibitem[Perfecto et~al.(2019)Perfecto, Vandermeer, and Wright]{Perfecto2016}
Ivette Perfecto, John Vandermeer, and Angus Wright.
\newblock \emph{Nature's matrix: linking agriculture, biodiversity conservation and food sovereignty}.
\newblock Routledge, 2019.

\bibitem[Phalan et~al.(2011)Phalan, Onial, Balmford, and Green]{Phalan2011}
Ben Phalan, Malvika Onial, Andrew Balmford, and Rhys~E Green.
\newblock Reconciling food production and biodiversity conservation: land sharing and land sparing compared.
\newblock \emph{science}, 333\penalty0 (6047):\penalty0 1289--1291, 2011.

\bibitem[Pierce and Nowak(1999)]{pierce1999}
Francis~J Pierce and Peter Nowak.
\newblock Aspects of precision agriculture.
\newblock \emph{Advances in agronomy}, 67:\penalty0 1--85, 1999.

\bibitem[Pingali(2012)]{Pingali2012}
Prabhu~L Pingali.
\newblock Green revolution: impacts, limits, and the path ahead.
\newblock \emph{Proceedings of the national academy of sciences}, 109\penalty0 (31):\penalty0 12302--12308, 2012.

\bibitem[Poore and Nemecek(2018)]{Poore2018}
Joseph Poore and Thomas Nemecek.
\newblock Reducing food’s environmental impacts through producers and consumers.
\newblock \emph{Science}, 360\penalty0 (6392):\penalty0 987--992, 2018.

\bibitem[Reganold and Wachter(2016)]{Reganold2016}
John~P Reganold and Jonathan~M Wachter.
\newblock Organic agriculture in the twenty-first century.
\newblock \emph{Nature plants}, 2\penalty0 (2):\penalty0 1--8, 2016.

\bibitem[Rey~Benayas et~al.(2007)Rey~Benayas, Martins, Nicolau, and Schulz]{rey2007abandonment}
Jos{\'e}~Mar{\'\i}a Rey~Benayas, Ana Martins, Jose~M Nicolau, and Jennifer~J Schulz.
\newblock Abandonment of agricultural land: an overview of drivers and consequences.
\newblock \emph{CABI Reviews}, \penalty0 (2007):\penalty0 14--pp, 2007.

\bibitem[Rigby and C{\'a}ceres(2001)]{rigby2001organic}
Dan Rigby and Daniel C{\'a}ceres.
\newblock Organic farming and the sustainability of agricultural systems.
\newblock \emph{Agricultural systems}, 68\penalty0 (1):\penalty0 21--40, 2001.

\bibitem[Rockstr{\"o}m et~al.(2009)Rockstr{\"o}m, Steffen, Noone, Persson, Chapin~III, Lambin, Lenton, Scheffer, Folke, Schellnhuber, et~al.]{Rockstrom2009}
Johan Rockstr{\"o}m, Will Steffen, Kevin Noone, {\AA}sa Persson, F~Stuart Chapin~III, Eric~F Lambin, Timothy~M Lenton, Marten Scheffer, Carl Folke, Hans~Joachim Schellnhuber, et~al.
\newblock A safe operating space for humanity.
\newblock \emph{nature}, 461\penalty0 (7263):\penalty0 472--475, 2009.

\bibitem[Rockstr{\"o}m et~al.(2020)Rockstr{\"o}m, Edenhofer, Gaertner, and DeClerck]{Rockstrom2020}
Johan Rockstr{\"o}m, Ottmar Edenhofer, Juliana Gaertner, and Fabrice DeClerck.
\newblock Planet-proofing the global food system.
\newblock \emph{Nature food}, 1\penalty0 (1):\penalty0 3--5, 2020.

\bibitem[Scheffer et~al.(2000)Scheffer, Brock, and Westley]{scheffer2000}
Marten Scheffer, William Brock, and Frances Westley.
\newblock Socioeconomic mechanisms preventing optimum use of ecosystem services: an interdisciplinary theoretical analysis.
\newblock \emph{Ecosystems}, 3\penalty0 (5):\penalty0 451--471, 2000.

\bibitem[Scheffer et~al.(2001)Scheffer, Carpenter, Foley, Folke, and Walker]{Scheffer2001}
Marten Scheffer, Steve Carpenter, Jonathan~A Foley, Carl Folke, and Brian Walker.
\newblock Catastrophic shifts in ecosystems.
\newblock \emph{Nature}, 413\penalty0 (6856):\penalty0 591--596, 2001.

\bibitem[Searchinger et~al.(2008)Searchinger, Heimlich, Houghton, Dong, Elobeid, Fabiosa, Tokgoz, Hayes, and Yu]{Searchinger2008}
Timothy Searchinger, Ralph Heimlich, Richard~A Houghton, Fengxia Dong, Amani Elobeid, Jacinto Fabiosa, Simla Tokgoz, Dermot Hayes, and Tun-Hsiang Yu.
\newblock Use of us croplands for biofuels increases greenhouse gases through emissions from land-use change.
\newblock \emph{Science}, 319\penalty0 (5867):\penalty0 1238--1240, 2008.

\bibitem[Searchinger et~al.(2019)Searchinger, Waite, Hanson, Ranganathan, Dumas, Rizet, Matthews, Klirs, Mallet, Adeniji, Blanchard, Boval, Cheung, Dann, Desjonqu\`eres, Dumas, Guyomard, Lecocq, Quinio, and Havlik]{Searchinger2019}
Timothy~D. Searchinger, Richard Waite, Craig Hanson, Janet Ranganathan, Patrice Dumas, Christophe Rizet, Evan Matthews, Christian Klirs, Patrick Mallet, Adedoyin Adeniji, Tatyana Blanchard, Michel Boval, Winnie Cheung, Louise Dann, Sarah Desjonqu\`eres, Pierre Dumas, Herve Guyomard, Francois Lecocq, Marion Quinio, and Petr Havlik.
\newblock Creating a sustainable food future.
\newblock \emph{World Resources Report}, 2019.
\newblock URL \url{https://www.wri.org/research/creating-sustainable-food-future}.
\newblock World Resources Institute.

\bibitem[Seufert et~al.(2012)Seufert, Ramankutty, and Foley]{Seufert2012}
Verena Seufert, Navin Ramankutty, and Jonathan~A Foley.
\newblock Comparing the yields of organic and conventional agriculture.
\newblock \emph{Nature}, 485\penalty0 (7397):\penalty0 229--232, 2012.

\bibitem[Smith and Gregory(2013)]{Smith2013}
Pete Smith and Peter~J Gregory.
\newblock Climate change and sustainable food production.
\newblock \emph{Proceedings of the nutrition society}, 72\penalty0 (1):\penalty0 21--28, 2013.

\bibitem[Soto et~al.(2019)Soto, Barnes, Balafoutis, Beck, S{\'a}nchez, Vangeyte, Fountas, Van~der Wal, Eory, and G{\'o}mez-Barbero]{soto2019}
Iria Soto, Andrew Barnes, Athanasios Balafoutis, Bert Beck, Berta S{\'a}nchez, Jurgen Vangeyte, Spyros Fountas, Tamme Van~der Wal, Vera Eory, and Manuel G{\'o}mez-Barbero.
\newblock \emph{The contribution of precision agriculture technologies to farm productivity and the mitigation of greenhouse gas emissions in the EU}.
\newblock Publications Office of the European Union Luxembourg, 2019.

\bibitem[Springmann et~al.(2016)Springmann, Godfray, Rayner, and Scarborough]{Springmann2016}
Marco Springmann, H.~Charles~J. Godfray, Mike Rayner, and Peter Scarborough.
\newblock Analysis and valuation of the health and climate change cobenefits of dietary change.
\newblock \emph{Proceedings of the National Academy of Sciences}, 113\penalty0 (15):\penalty0 4146--4151, 2016.

\bibitem[Springmann et~al.(2018)Springmann, Clark, Mason-D’Croz, Wiebe, Bodirsky, Lassaletta, De~Vries, Vermeulen, Herrero, Carlson, et~al.]{Springmann2018}
Marco Springmann, Michael Clark, Daniel Mason-D’Croz, Keith Wiebe, Benjamin~Leon Bodirsky, Luis Lassaletta, Wim De~Vries, Sonja~J Vermeulen, Mario Herrero, Kimberly~M Carlson, et~al.
\newblock Options for keeping the food system within environmental limits.
\newblock \emph{Nature}, 562\penalty0 (7728):\penalty0 519--525, 2018.

\bibitem[Starbird et~al.(2016)Starbird, Norton, and Marcus]{starbird2016investing}
Ellen Starbird, Maureen Norton, and Rachel Marcus.
\newblock Investing in family planning: key to achieving the sustainable development goals.
\newblock \emph{Global health: science and practice}, 4\penalty0 (2):\penalty0 191--210, 2016.

\bibitem[Stehfest et~al.(2014)Stehfest, van Vuuren, Bouwman, Kram, et~al.]{Stehfest2014}
Elke Stehfest, Detlef van Vuuren, Lex Bouwman, Tom Kram, et~al.
\newblock \emph{Integrated assessment of global environmental change with IMAGE 3.0: Model description and policy applications}.
\newblock Netherlands Environmental Assessment Agency (PBL), 2014.

\bibitem[Tilman et~al.(2011)Tilman, Balzer, Hill, and Befort]{Tilman2011}
David Tilman, Christian Balzer, Jason Hill, and Belinda~L Befort.
\newblock Global food demand and the sustainable intensification of agriculture.
\newblock \emph{Proceedings of the national academy of sciences}, 108\penalty0 (50):\penalty0 20260--20264, 2011.

\bibitem[Verburg et~al.(2011)Verburg, Neumann, and Nol]{Verburg2013}
Peter~H Verburg, Kathleen Neumann, and Linda Nol.
\newblock Challenges in using land use and land cover data for global change studies.
\newblock \emph{Global change biology}, 17\penalty0 (2):\penalty0 974--989, 2011.

\bibitem[Wang et~al.(2023)Wang, Wang, Chen, and Zhao]{wang2023vertical}
Gangyi Wang, Jingjing Wang, Siyu Chen, and Chang{\'{}}~e Zhao.
\newblock Vertical integration selection of chinese pig industry chain under african swine fever-from the perspective of stable pig supply.
\newblock \emph{Plos one}, 18\penalty0 (2):\penalty0 e0280626, 2023.

\bibitem[Watson et~al.(2018)Watson, Evans, Venter, Williams, Tulloch, Stewart, Thompson, Ray, Murray, Salazar, et~al.]{watson2018exceptional}
James~EM Watson, Tom Evans, Oscar Venter, Brooke Williams, Ayesha Tulloch, Claire Stewart, Ian Thompson, Justina~C Ray, Kris Murray, Alvaro Salazar, et~al.
\newblock The exceptional value of intact forest ecosystems.
\newblock \emph{Nature ecology \& evolution}, 2\penalty0 (4):\penalty0 599--610, 2018.

\bibitem[Willett et~al.(2019)Willett, Rockstr{\"o}m, Loken, Springmann, Lang, Vermeulen, Garnett, Tilman, DeClerck, Wood, et~al.]{Willett2019}
Walter Willett, Johan Rockstr{\"o}m, Brent Loken, Marco Springmann, Tim Lang, Sonja Vermeulen, Tara Garnett, David Tilman, Fabrice DeClerck, Amanda Wood, et~al.
\newblock Food in the anthropocene: the eat--lancet commission on healthy diets from sustainable food systems.
\newblock \emph{The lancet}, 393\penalty0 (10170):\penalty0 447--492, 2019.

\bibitem[Zabel et~al.(2019)Zabel, Delzeit, Schneider, Seppelt, Mauser, and V{\'a}clav{\'\i}k]{vanDijk2020}
Florian Zabel, Ruth Delzeit, Julia~M Schneider, Ralf Seppelt, Wolfram Mauser, and Tom{\'a}{\v{s}} V{\'a}clav{\'\i}k.
\newblock Global impacts of future cropland expansion and intensification on agricultural markets and biodiversity.
\newblock \emph{Nature communications}, 10\penalty0 (1):\penalty0 2844, 2019.

\bibitem[Zeide(1991)]{zeide1991fractal}
Boris Zeide.
\newblock Fractal geometry in forestry applications.
\newblock \emph{Forest Ecology and Management}, 46\penalty0 (3-4):\penalty0 179--188, 1991.

\bibitem[zu~Ermgassen et~al.(2024)zu~Ermgassen, Renier, Garcia, Carvalho, and Meyfroidt]{zu2024sustainable}
Erasmus~KHJ zu~Ermgassen, C{\'e}cile Renier, Andrea Garcia, Tom{\'a}s Carvalho, and Patrick Meyfroidt.
\newblock Sustainable commodity sourcing requires measuring and governing land use change at multiple scales.
\newblock \emph{Conservation Letters}, 17\penalty0 (3):\penalty0 e13016, 2024.

\end{thebibliography}

\clearpage

\appendix
\section*{Appendix}
\setcounter{section}{0}
\renewcommand{\thesection}{A}

\subsection{Initialization}
\label{app:landscape_init}

The system initialization proceeds as follows:

\begin{itemize}
    \item The model employs a $100\times100$ grid to represent the landscape, initially allocating 50\% of the area to forest land. This proportion reflects estimates for the year 1960 based on FAO data~\cite{faostructuraldata}. To spatially distribute this fraction, a fractal field with a characteristic correlation length $\xi$ is generated. Previous research suggests that fractal patterns realistically capture empirical land use distributions~\cite{milne1991utility, zeide1991fractal}. The value for $\xi$, chosen to produce a visually realistic arrangement, is reported in Table~\ref{tab:params}. After allocating forested cells, the remaining landscape is randomly assigned to pasture (35\%) and cropland (15\%).
    
    \item Next, each cell receives specific degradation and restoration rates. These are assigned using the concept of lifespan, which aligns well with the properties of exponential distributions. This approach has been motivated by soil science literature, where the exponential lifetime of soil layers is tied to degradation and erosion processes~\cite{evans2020soil}. Following results this stream of research, each parameter $\theta_{i, \cdot}$ is sampled as:
    \begin{equation*}
        \theta_{i, \cdot} = \frac{1}{\mathcal{N}(\mu_{\cdot}, \sigma_{\cdot})},
    \end{equation*}
    where $\cdot$ stands for natural ($\mathrm{n}$), pasture ($\mathrm{m}$), or cropland ($\mathrm{c}$), and $\mathcal{N}$ denotes a lognormal distribution. The adoption of a lognormal distribution ensures the statistical characteristics of the rates are consistent with those reported by Evans et al.~\cite{evans2020soil}.

    \item Initial real demand for both crops and meat is computed following Eq.~\ref{eq:calories_demand} and Eq.~\ref{eq:meat_demand}. To translate this demand into the model’s internal units, we normalize by assigning constants $\omega_\cdot$ such that the initial yield for both livestock and crops is set to $q_{\cdot,0} = 1$. Assuming that at the starting point total production matches demand, we have $Q_{\mathrm{c},0}=1500=D_{\mathrm{c},0}$ and $Q_{\mathrm{m},0}=3500=D_{\mathrm{m},0}$. Since this step only rescales the reference point, the overall system dynamics are unaffected.

    \item Intensification factors are then initialized to have $q_{\cdot,0}=1$ in the initial year, which requires livestock density, mechanization, chemical input intensity, and average ecosystem integrity to all be set to one at initialization. In contrast, we assign the initial technology level a small value ($T_0=0.001$), reflecting the limited state of agricultural innovation prior to the Green Revolution. While this may result in yields slightly above one, the difference is negligible for trend analysis.

    \item With all state variables initialized, the model is ready to proceed with the simulation.
\end{itemize}

\subsection{Calibration procedure and parameter values} 
\label{app:calibration_parameters}

Model parameters were categorized into four distinct groups (see Table~\ref{tab:params}): (i) \textit{measured} parameters (\textcolor{measuredColor}{M}), which are directly fitted from empirical datasets; (ii) \textit{estimated} parameters (\textcolor{estimatedColor}{E}), sourced from values reported in the literature; (iii) \textit{calibrated} parameters (\textcolor{calibratedColor}{C}), which are adjusted during model calibration to achieve the best agreement between simulated and observed data; and (iv) \textit{fixed} parameters (\textcolor{fixedColor}{F}), which are manually specified due to the lack of real-world analogs or because of model simplifications.

\begin{table*}[h!]
    \centering
    \begin{tabular}{lcllc}
        \multicolumn{5}{c}{~} \\ \midrule
        \toprule
        Section & Notation & Description & Value & Group \\ 
        \midrule 
        Landscape &  & Grid size & 100x100 &  \textcolor{fixedColor}{F} \\
        Initialization &  & Natural land share & 50\% &  \textcolor{measuredColor}{M} \\
         &  & Cropland share & 15\% &  \textcolor{measuredColor}{M} \\
         &  & Pasture land share & 35\% &  \textcolor{measuredColor}{M} \\
         & $\xi$ & Landscape fragmentation  & 10 & \textcolor{fixedColor}{F} \\
         & $\mu_{\mathrm{c}}$ & Crop degradation distribution shift & 1 & \textcolor{estimatedColor}{E} \\
         & $\mu_{\mathrm{m}}$ & Pasture degradation distribution shift & 1 & \textcolor{estimatedColor}{E} \\
         & $\sigma_{\mathrm{c}}$ & Crop degradation distribution scale & 10000 & \textcolor{estimatedColor}{E} \\
         & $\sigma_{\mathrm{m}}$ & Pasture degradation distribution scale & 5000 & \textcolor{estimatedColor}{E} \\
         & $\mu_{\mathrm{n}}$ & Nature restoration distribution shift & 1 & \textcolor{estimatedColor}{E} \\
         & $\sigma_{\mathrm{n}}$ & Pasture degradation distribution scale & 1000 & \textcolor{estimatedColor}{E} \\
         \cdashline{1-5}
        Demand & $a$ & Calories demand calculation & -138.2 &  \textcolor{measuredColor}{M} \\
         & &  \footnotesize{$[ \mathrm{kcal}\cdot \text{year}^{-1} \cdot \text{per capita}]$}& & \\
         & $b$ & Calories demand calculation & 744.4 &  \textcolor{measuredColor}{M} \\
         & &  \footnotesize{$[ \mathrm{kcal}\cdot \text{\$}^{-1} \cdot \text{per capita}]$}& & \\
         & $c$ & Baseline demand for animal products & 210 & \textcolor{measuredColor}{M} \\
         & &  \footnotesize{$[ \mathrm{Kg}\cdot \text{year}^{-1} \cdot \text{per capita}]$}& & \\
         & $r$ & Conversion demand for animal products to calories & 3000 & \textcolor{estimatedColor}{E} \\
         & &  \footnotesize{$[ \mathrm{kcal}\cdot \text{Kg}^{-1}]$}& & \\
         & $d$ & Income elasticity of meat demand & 0.65 & \textcolor{measuredColor}{M} \\
         & $\alpha_m$ & Feedback strength for meat demand & 0.5 & \textcolor{calibratedColor}{C} \\
         & $\alpha_c$ & Feedback strength for crop demand & 0.1 & \textcolor{calibratedColor}{C} \\
         \cdashline{1-5}
        Production & $k$ & Yield elasticity to chemical inputs & 0.2 & \textcolor{estimatedColor}{E} \\
         & $f$ & Yield elasticity to mechanization & 0.5 & \textcolor{estimatedColor}{E} \\
         & $h$ & Sensitivity of meat yield to intensification & 0.95 & \textcolor{estimatedColor}{E} \\
         & $\Lambda_{\max}$ & Maximum livestock density (organic) & 3 & \textcolor{estimatedColor}{E} \\
         & $\beta$ & Speed of mechanization adjustment & 0.95 & \textcolor{calibratedColor}{C} \\
         & $\delta$ & Speed of chemical input adjustment & 1.1 & \textcolor{calibratedColor}{C} \\
         & $\gamma$ & Speed of livestock density adjustment & 0.95 & \textcolor{calibratedColor}{C} \\
         & $\nu$ & Technological progress rate & 0.1 &  \textcolor{estimatedColor}{E} \\
         & $T_{\max}$ & Technological ceiling & 0.2 &  \textcolor{estimatedColor}{E} \\
         \cdashline{1-5}
        Landscape  & $\varphi$ & Baseline land expansion rate (crop/pasture) & 3e-4 & \textcolor{calibratedColor}{C} \\
        dynamics & $\zeta^+_{\mathrm{c}}$ & Expansion rate (cropland) & 130 & \textcolor{calibratedColor}{C} \\
         & $\zeta^+_{\mathrm{m}}$ & Expansion rate (pasture land) & 500 &\textcolor{calibratedColor}{C} \\
         & $\zeta^-_{\mathrm{c}}$ & Contraction rate (cropland) & 120 & \textcolor{calibratedColor}{C} \\
         & $\zeta^-_{\mathrm{m}}$ & Contraction rate (pasture land) & 500 & \textcolor{calibratedColor}{C} \\
         & $\varepsilon_{\max}$ & Maximum natural ecosystem integrity & 2 &  \textcolor{fixedColor}{F} \\
         & $p$ & Diminishing return for ecosystem service  & 0.25& \textcolor{estimatedColor}{E} \\
         \midrule
    \end{tabular}
    \caption{Parameters of the model. Group codes can be: measured (M), estimated (E), calibrated (C), or fixed (F).}
    \label{tab:params}
\end{table*}

\textit{Measured} parameters (\textcolor{measuredColor}{M}) are defined as those whose values are determined using empirical data corresponding directly to the model functions they influence. This group mainly encompasses parameters governing demand and the initial state of the landscape, where reliable datasets are readily available. Parameter values were estimated by fitting model outputs to observed data from FAO\cite{faostructuraldata, FAO2024_meat}, using a least squares minimization approach to ensure alignment with real-world patterns. It should be emphasized that the aim of this calibration was not to achieve exact reproduction of the full distribution of empirical curves, but rather to capture the essential trends relevant to the aims of this study. Within this context, all fitted functions yield a robust agreement with the empirical data, adequately representing the characteristic patterns for each parameter considered (see Fig.~\ref{fig:calibration}).

\textit{Estimated} parameters (\textcolor{estimatedColor}{E}) were assigned values drawn from empirical studies and widely reported literature benchmarks. This category chiefly includes ecological parameters such as rates of degradation and restoration, for which precise values are not directly measurable within the model context. As described in Section~\ref{app:landscape_init}, these rates were informed by global analyses and meta-analyses covering soil degradation and forest recovery dynamics\cite{evans2020soil, garcia2015meta, crouzeilles2016global}. For parameters associated with the yield function, values were chosen based on extensive agricultural economics and ecological economics literature\cite{evenson2009handbook, cassman1999ecological, Mitcherliche_prod_func}. While reported values do vary between studies and there is no universal consensus—often due to differences in scale, crop types, or methodological approaches—our selections are consistent with established ranges. Importantly, the chosen parameter values are supported by the model's ability to reproduce historical patterns of yield growth observed in real-world data, lending further confidence to their plausibility within the framework of this study.

Parameters classified as \textit{calibrated} (\textcolor{calibratedColor}{C}) are primarily associated with the model’s adjustment and feedback mechanisms. These parameters cannot be directly inferred from empirical data as they are fundamentally linked to internal model behavior—that is, they govern logical or process-based linkages for which data are sparse or context-specific. Accordingly, they were tuned manually to best reproduce the broad patterns present in observed system dynamics, using an iterative, trial-and-error approach. The goal here, as elsewhere, was not to exactly fit every empirical curve but to recapitulate the essential dynamic behavior pertinent to our research questions. This calibration approach ensures the model captures the relevant system variability while maintaining enough flexibility to allow exploration of alternative scenarios within the stylized earth system structure.

\end{document}